%% file: main.tex
\documentclass{article}

\usepackage[T1]{fontenc}
\usepackage[utf8]{inputenc}
\usepackage{lmodern}
\usepackage[english]{babel}
\usepackage[autostyle]{csquotes}

\usepackage{fullpage}
\usepackage[ocgcolorlinks,pdfusetitle]{hyperref}
\usepackage{doi}
\usepackage{xcolor}
\usepackage{xspace}
\usepackage{graphicx}
\usepackage{tikz}
\usetikzlibrary{positioning, shapes.geometric, arrows.meta, calc}

\usepackage{natbib}
\usepackage{subcaption}
\usepackage{booktabs}  % (\toprule, \midrule, \bottomrule)
\usepackage{tabularx}  
\usepackage{caption}   
\usepackage{pdflscape}
\usepackage{makecell}

\input{math}

\definecolor{mydarkblue}{rgb}{0,0.08,0.45}
\hypersetup{
colorlinks=true,
urlcolor=mydarkblue,
linkcolor=mydarkblue,
citecolor=mydarkblue,
}

\begin{document}
% \title{Generative AI for Protein Dynamics Modeling} % old
\title{Learning Structure, Energy, and Dynamics: A Survey of Artificial Intelligence for Protein Dynamics}
\author{
Haocheng Tang$^{1,2,*,\ddagger}$\quad
Liang Shi$^{3,*}$\quad
Ya-Shi Zhang$^{1,4,*}$\quad
Xixian Liu$^{1,4}$ \\
Jian Tang$^{1,5,6,\dagger}$\quad
Jiarui Lu$^{1,4,\dagger}$
\\[6pt]
$^1$Mila -- Quebec AI Institute\quad
$^2$University of Pittsburgh\quad
$^3$Peking University\\
$^4$Universit\'{e} de Montr\'{e}al\quad
$^5$HEC Montr\'{e}al\quad
$^6$CIFAR AI Chair
}

\date{\today}

\maketitle

\begingroup
\renewcommand{\thefootnote}{}
\footnotetext{$^*$Equal contribution.}
\footnotetext{$^\dagger$Correspondence to \texttt{jiarui.lu@umontreal.ca, jian.tang@hec.ca}}
\footnotetext{$^\ddagger$Work conducted during an internship at Mila}
\endgroup

% ===================================
% main text here 
% ===================================
\begin{abstract}

Protein dynamics underlie many biological functions, yet remain difficult to characterize due to the high computational cost of molecular dynamics simulations and the scarcity of dynamic structural data. This survey reviews recent advances in artificial intelligence for protein dynamics from three perspectives: learning from structural ensembles and trajectories, learning from physical energy signals, and learning to accelerate molecular simulations. We summarize representative methods for conformation ensemble generation, trajectory generation, Boltzmann generators, physics-aware adaptation, machine learning potentials, coarse-grained modeling, and collective variable discovery. We further discuss available datasets and key open challenges, such as scalability, thermodynamic consistency, kinetic fidelity, and integration with experimental constraints.

\end{abstract}

\section{Introduction}
% \jr{\textbf{Note for everyone}:  When adding bibtex to the bib file \texttt{refs.bib}, use journal's or proceedings' bibtex if already published; avoid arxiv bibtex except for unpublished works. }
\iffalse
\textbf{The current outline for the main text:}
\begin{enumerate}
    \item Section 2; Learning to generate from structures (conformation generation, transition operator, trajectory generation)
    \item Section 3; Learning to generate from energy (BG, energy-based training)
    \item Section 4; Learning for MD simulations (ML FFs, ML CVs)
    \item MISC: Training Data; Evaluation metrics; Discussion
\end{enumerate}
\fi

Proteins are highly dynamic macromolecules whose biological functions---ranging from enzymatic catalysis and signal transduction to allosteric regulation---are intrinsically tied to their conformational flexibility. While determining a single, static structure provides critical insights, a comprehensive understanding of biological mechanisms demands the characterization of the full conformational ensemble and the dynamic transition pathways between different functional states. Consequently, modeling protein dynamics is a central challenge in structural biology, biophysics, and structure-based drug discovery.

Traditionally, Molecular Dynamics (MD) simulation has been the workhorse for investigating protein dynamics at atomic resolution. MD integrates Newton's equations of motion using empirical or quantum mechanical force fields to simulate the continuous evolution of molecular systems. However, a fundamental limitation of MD is its immense computational cost. Capturing high-frequency atomic vibrations necessitates femtosecond-scale integration time steps, which makes sampling long-timescale biological events---such as protein folding, large-scale conformational rearrangements, and complex ligand binding---computationally prohibitive for most realistic systems. 

In recent years, the landscape of structural biology has been revolutionized by Deep Learning (DL) and Generative Artificial Intelligence (GenAI). Following the unprecedented success of static protein structure prediction models like AlphaFold and ESMFold, the research frontier has rapidly expanded toward modeling protein dynamics. Generative AI models, including diffusion models, flow matching, and autoregressive models, are now being tailored to sample diverse conformational ensembles, interpolate transition paths, and even generate complete dynamic trajectories directly. Furthermore, machine learning techniques are increasingly tightly integrated with physical principles, offering new ways to enhance, surrogate, or bypass traditional MD simulations.

In this survey, we provide a comprehensive review of the rapidly evolving field of Generative AI and Machine Learning for protein dynamics modeling. We categorize the existing literature into three primary perspectives based on their underlying methodologies and training signals:
\begin{itemize}
    \item \textbf{Learning from Structural Data} (Section~2): We review generative models trained directly on discrete structural ensembles or continuous MD trajectories. This includes methods for sampling equilibrium conformation ensembles, generating sequence-conditional conformers, and developing autoregressive or flow-based models for continuous trajectory generation.
    \item \textbf{Learning from Energy Signals} (Section~3): We discuss methods that rely on physical energy functions, rather than solely on empirical data, to learn the underlying Boltzmann distribution. Models in this category include Boltzmann Generators, which are trained via energy-based objectives or annealed importance sampling, as well as physics-aware parameter fine-tuning and guidance of pre-trained generative models.
    \item \textbf{Learning for MD Simulations} (Section~4): We explore the integration of machine learning to accelerate and refine classical simulations. This encompasses Machine Learning Potentials (MLPs) for high-fidelity force fields, Machine Learning Coarse-Grained (CG) models for dimensionality reduction, and ML-derived Collective Variables (CVs) for enhanced sampling.
\end{itemize}

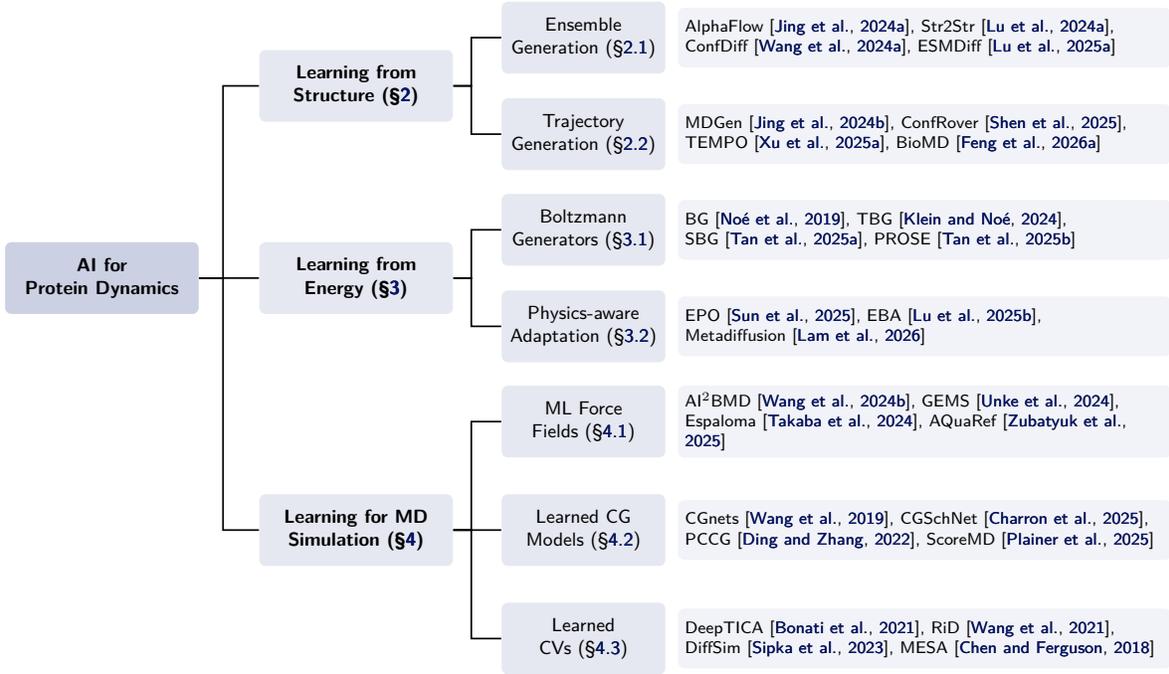
\begin{figure}[htbp!]
    \centering
    \resizebox{0.95\textwidth}{!}{

    \begin{tikzpicture}[
        box/.style={rectangle, rounded corners=3pt, minimum height=1.2cm, text width=3.0cm, align=center, fill=mydarkblue!10, text=black, font=\small\sffamily},
        mainbox/.style={box, text width=3.0cm, font=\small\bfseries\sffamily},
        smallbox/.style={box, text width=2.5cm, font=\small\sffamily},
        listbox/.style={rectangle, rounded corners=3pt, minimum height=1.0cm, text width=8cm, align=left, fill=mydarkblue!5, text=black, font=\footnotesize\sffamily},
        line/.style={draw, thick, -}
    ]

    % Root
    \node[mainbox, fill=mydarkblue!20] (root) at (0, 0) {AI for\\Protein Dynamics};

    % Level 1
    \node[mainbox] (l2) [right=1cm of root] {Learning from\\Energy (\S\ref{sec:energy_signals})};
    \node[mainbox] (l1) [above=2cm of l2] {Learning from\\Structure (\S\ref{sec:structural_data})};
    \node[mainbox] (l3) [below=3cm of l2] {Learning for MD\\Simulation (\S\ref{sec:learn_mds})};

    \draw[line] (root.east) -- +(0.4,0) |- (l1.west);
    \draw[line] (root.east) -- (l2.west);
    \draw[line] (root.east) -- +(0.4,0) |- (l3.west);

% Level 2 for l1
    \node[smallbox] (l11) [above right=0.2cm and 0.8cm of l1.east] {Ensemble\\Generation (\S\ref{subsec:ensemble})};
    \node[smallbox] (l12) [below right=0.2cm and 0.8cm of l1.east] {Trajectory\\Generation (\S\ref{subsec:trajectory})};

    \draw[line] (l1.east) -- +(0.3,0) |- (l11.west);
    \draw[line] (l1.east) -- +(0.3,0) |- (l12.west);

    % Level 2 for l2
    \node[smallbox] (l21) [above right=0.2cm and 0.8cm of l2.east] {Boltzmann\\Generators (\S\ref{subsec:boltzmann})};
    \node[smallbox] (l22) [below right=0.2cm and 0.8cm of l2.east] {Physics-aware\\Adaptation (\S\ref{subsec:physics_adapt})};

    \draw[line] (l2.east) -- +(0.3,0) |- (l21.west);
    \draw[line] (l2.east) -- +(0.3,0) |- (l22.west);

    % Level 2 for l3
    \node[smallbox] (l32) [right=0.8cm of l3.east] {Learned CG\\Models (\S\ref{subsec:mlcg})};
    \node[smallbox] (l31) [above=0.6cm of l32] {ML Force\\Fields (\S\ref{subsec:mlp})};
    \node[smallbox] (l33) [below=0.6cm of l32] {Learned\\CVs (\S\ref{subsec:mlcv})};

    \draw[line] (l3.east) -- +(0.3,0) |- (l31.west);
    \draw[line] (l3.east) -- (l32.west);
    \draw[line] (l3.east) -- +(0.3,0) |- (l33.west);

    % Level 3
    \node[listbox] (r11) [right=0.2cm of l11] {AlphaFlow~\citep{jing2024alphaflow}, Str2Str~\citep{lu2024strstr}, \\ConfDiff~\citep{wang2024confdiff}, ESMDiff~\citep{lu2025slm}};
    \node[listbox] (r12) [right=0.2cm of l12] {MDGen~\citep{jing2024mdgen}, ConfRover~\citep{shen2025confrover}, \\TEMPO~\citep{xu2025tempo}, BioMD~\citep{feng2026biomd}};

    \node[listbox] (r21) [right=0.2cm of l21] {BG~\citep{noe2019bg}, TBG~\citep{klein2024tbg}, \\SBG~\citep{tan2025sbg}, PROSE~\citep{tan2025prose}};
    \node[listbox] (r22) [right=0.2cm of l22] {EPO~\citep{sun2025epo}, EBA~\citep{lu2025eba}, \\Metadiffusion~\citep{lam2026metadiffusion}};

    \node[listbox] (r31) [right=0.2cm of l31] {AI$^2$BMD~\citep{wang2024ai2bmd}, GEMS~\citep{unke2024gems}, \\Espaloma~\citep{takaba2024espaloma}, AQuaRef~\citep{zubatyuk2025aquaref}};
    \node[listbox] (r32) [right=0.2cm of l32] {CGnets~\citep{wang2019cgnet}, CGSchNet~\citep{charron2025cgschnet}, \\PCCG~\citep{ding2022pccg}, ScoreMD~\citep{plainer2025consistent}};
    \node[listbox] (r33) [right=0.2cm of l33] {DeepTICA~\citep{luigi2021deeptica}, RiD~\citep{wang2021rid}, \\DiffSim~\citep{sipka2023diffsim}, MESA~\citep{chen2018mesa}};

    \end{tikzpicture}
    }
    \caption{Taxonomy of methods and representative works for each direction.}
    \label{fig:taxonomy}
\end{figure}

Finally, we catalog publicly available datasets that are instrumental for training and benchmarking these models (Section~5), and conclude with a discussion of current challenges and promising future directions in the field (Section~6). 

\begin{figure}[htbp!]
    \centering
    \includegraphics[width=0.98\linewidth]{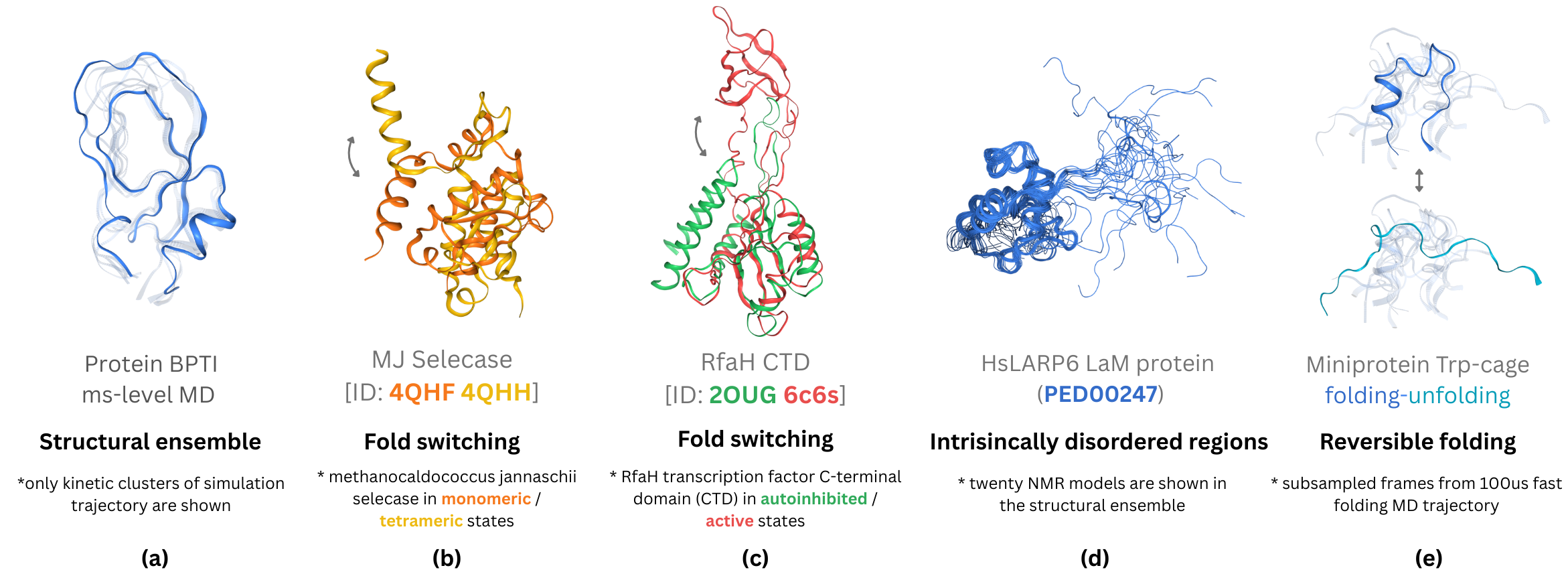}
    \caption{Examples of biomolecular conformational dynamics. This figure illustrates diverse dynamic phenomena observed via computational or experimental methods: \textbf{(a)} the structural ensemble of BPTI (PDB: 5PTI), represented by kinetic clusters from millisecond-level molecular dynamics~\citep{shaw2010atomic}; \textbf{(b)} observed fold switching in MJ selecase (PDB: 4QHF/4QHH) between monomeric and tetrameric states; \textbf{(c)} the autoinhibited ($\alpha$-helical hairpin) and the active ($\beta$-barrel) states of the RfaH transcription factor C-terminal domain (PDB: 2OUG/6C6S); 
    \textbf{(d)} intrinsically disordered regions (IDR) in the HsLARP6 LaM protein (NMR ensemble, PED00247); and \textbf{(e)} snapshots of reversible folding reaction of the Trp-cage (PDB: 2JOF) sampled from a $100\ \mu\text{s}$ trajectory~\citep{lindorff2011fast}.}
    \label{fig:intro-example}
\end{figure}

% \jr{to be deleted}
% \paragraph{Notations.} We denote a protein structure as $\rvx$ and the corresponding protein sequence as $\rvy$. A sequence of protein structures (protein conformation trajectory) is denoted by $\rmX$, and $\rvv$ typically refers to velocity in MD simulation. We use $p(\rvx)$ to denote a distribution over protein structures (and variants including Boltzmann distribution, data distribution, etc.). $E(\rvx)$ represents a force field or energy function.

% formatted table for summarizing methods
% method name; structure representation; training data; method/training objective; test systems.

\section{Learning from Structural Data}\label{sec:structural_data}

% \todo{@liang + @jiarui}

% method name; structure representation; training data; method/training objective; test systems; input structure or not

This section reviews the generative methods for conformation ensemble generation and trajectory generation. Figure~\ref{fig:section2} provides an overview of this section.

\subsection{Conformation ensemble generation}\label{subsec:ensemble}

\paragraph{Problem Formulation.} The task of protein ensemble generation involves modeling the distribution $p(\rvx)$ over protein structures $\rvx$ given a protein sequence $\rvy$. In a physical context, this target distribution corresponds to the Boltzmann distribution $p(\rvx) \propto \exp(-E(\rvx)/k_B T)$ determined by the energy function $E(\rvx)$. Generative models seek to approximate (emulate) this distribution, often leveraging datasets of crystal structures and simulation conformations.

\paragraph{Motivation} The biological function of a protein is governed not by a single static structure but by an ensemble of interconverting conformations, encompassing ordered fluctuations, coordinated large-scale motions, and transitions between distinct functional states. Accurately modeling this equilibrium distribution $p(\rvx)$ is therefore fundamental to understanding mechanistic biology and guiding molecular design. The gold-standard computational method, Molecular Dynamics (MD) simulation, generates physically rigorous trajectories by integrating Newton's equations of motion. However, achieving sufficient sampling to cross high energy barriers and converge the Boltzmann distribution is often prohibitively expensive due to the femtosecond-scale time steps required, limiting its practical utility for exploring large-scale conformational landscapes. 

Deep learning has revolutionized single-state protein structure prediction. This success inspires a natural extension: leveraging deep models to directly approximate the distribution of accessible structures. Initial heuristic approaches, such as perturbing multiple sequence alignments (MSAs) as input to AlphaFold2~\citep{del2022msasub, wayment2024afcluster}, demonstrated the potential to reveal alternative conformations but suffer from critical limitations~\citep{jing2024alphaflow, wang2024confdiff}: (i) they operate only at inference time, preventing direct training on ensembles of known conformations; (ii) the generated conformational diversity is often limited; (iii) they are incompatible with emerging MSA-free predictors like ESMFold that leverage protein language models; and (iv) they provide no theoretical guarantee that the sampled distribution relates to a physically meaningful Boltzmann distribution. 

To overcome these constraints, a new paradigm employs explicit \textit{generative models} to learn and sample from $p(\rvx|\rvy)$. This framework enables direct training on diverse structural ensembles, offers the potential for efficient and diverse sampling, and provides a principled statistical foundation for emulating equilibrium protein dynamics.

\begin{table}[htbp!]
    \centering
    \scriptsize
    \caption{Overview of related methods of Learning from Structural Data. \textit{Protein Rep.} indicates the protein structural representation.}
    \label{tab:lit_review_methods}
    \begin{tabularx}{\textwidth}{@{}l l l l l@{}}
        \toprule
        \textbf{Method} & \textbf{Protein Rep.} & \textbf{Training Data} & \textbf{Method} & \textbf{Test Systems} \\
        \midrule
  
AlphaFlow~\citep{jing2024alphaflow}         & All-atom       & ATLAS, PDB                & Flow Matching & ATLAS, IDP, PDB           \\
AlphaFolding~\citep{cheng2025alphafolding}  & All-atom       & ATLAS, PDB                & Diffusion     & ATLAS, Fast Folders       \\
AlphaPPImd~\citep{wang2024alphappimd}       & Backbone       & Barnase-Barstar MD        & Latent LM     & Protein-protein Complexes                 \\
AnewSampling~\citep{wang2026learning}       & All-atom       & AnewSampling-DB            & Diffusion   & \makecell[l]{held-out PDB systems\\JACS \& Merck\\in-house MD} \\
aSAM~\citep{janson2025asam}                 & All-atom       & ATLAS, mdCATH             & Diffusion     & \makecell[l]{ATLAS, mdCATH, \\Fast Folders} \\
ATMOS~\citep{shi2026atmos}                  & All-atom       & \makecell[l]{PDB, mdCATH,\\MISATO}       & SSM + Diffusion     & mdCATH, MISATO       \\
BioEmu~\citep{lewis2025bioemu}              & Backbone       & \makecell[l]{AFDB, public/in-house\\MD}  & Diffusion     & \makecell[l]{Complexes, Fast Folders,\\IDP} \\
BioKinema~\citep{feng2026physically}         & All-atom       & DD-13M, MISATO            & Flow Matching & Complexes                 \\
BioMD~\citep{feng2026biomd}                 & All-atom       & DD-13M, MISATO            & Flow Matching & Complexes                 \\
Boltz2~\citep{passaro2025boltz}             & All-atom       & PDB, ATLAS, mdCATH        & Diffusion & PDB, ATLAS, mdCATH   \\
ConfDiff~\citep{wang2024confdiff}           & Backbone       & PDB                       & Diffusion     & BPTI, Fast Folders        \\
ConfRover~\citep{shen2025confrover}         & All-atom       & ATLAS                     & AR + Diffusion     & ATLAS                     \\
DeepJump~\citep{costa2025deepjump}          & All-atom       & mdCATH                    & Flow matching & Fast-folding MD      \\
DiG~\citep{zheng2024dig}                    & Backbone       & GPCRmd, PDB, PDBbind      & Diffusion     & PDB, IDP            \\
DynaFold~\citep{fan2025dynafold}            & All-atom       & AFDB, ATLAS, PDB          & Diffusion     & ATLAS, Fast Folders       \\
DynamicBind~\citep{lu2024dynamicbind}       & All-atom       & PDBbind                   & Diffusion     & Complexes                 \\
EquiJump~\citep{costa2024equijump}          & All-atom       & Fast-folding MD           & \makecell[l]{Stochastic\\Interpolants}   &  Fast-folding MD \\
ESMDiff~\citep{lu2025slm}                   & Backbone       & PDB                       & Latent LM     & PDB, BPTI, IDP       \\
% EBA~\citep{lu2025eba}                       & All-atom       & PDB, ATLAS               & Diffusion     & ATLAS     \\
F$^3$low~\citep{li2024f}                    & Backbone       & Fast-folding MD           & Flow-matching & Fast-folding MD       \\
GLDP~\citep{sengar2026beyond}               & All-atom       & \makecell[l]{Peptides, ATLAS,\\public MD} & Encoder-Decoder & \makecell[l]{Peptides, ATLAS,\\public MD} \\
IDPFold~\citep{zhu2025accurate}             & Backbone       & IDRome, PDB               & Diffusion     & IDP                       \\
idpGAN~\citep{janson2023idpgan}             & C$\alpha$      & CG MD                     & GAN           & IDP                       \\
ITO~\citep{schreiner2023ito}                & C$\alpha$      & Peptides/Fast-folding     & diffusion     & Peptides/Fast-folding     \\
MARS-FM~\citep{kapusniak2025mars}           & All-atom       & mdCATH, Peptides MD       & Flow Matching & Peptides, mdCATH            \\
MD-LLM-1~\citep{murtada2025mdllm1} & Latent & Mad2/T4 MD               & Latent LM     & Mad2, T4 Lysozyme         \\
MDGen~\citep{jing2024mdgen}                 & All-atom       & ATLAS, Peptides MD        & Flow Matching & ATLAS, Peptides           \\
P2DFlow~\citep{jin2025p2dflow}              & All-atom       & ATLAS                     & Flow Matching & ATLAS, Complexes          \\
PLACER~\citep{ivan2025placer}               & All-atom    & CSD, PDB                  & Denoising     & Complexes, Small Mol.     \\
PTraj-Diff~\citep{xu2025ptraj-diff}         & Backbone & 1HPV/1BRS MD                    & Diffusion     & Complexes                 \\
PVB~\citep{yu2026unified}                   & All-atom       & \makecell[l]{small molecules MD, PDB,\\PDBBind, ATLAS, \\mdCATH, MISATO} & Encoder-Decoder & \makecell[l]{ATLAS, mdCATH,\\MISATO} \\
ScooBDoob~\citep{zhang2025scoobdoob}        & CV         & Peptides MD        & Bridge matching  & Peptides MD        \\
% ScoreMD is discussed in Section 4 (ML CG Models)
% ScoreMD~\citep{plainer2025consistent}       & Backbone/C$\alpha$ & Fast-folding MD       & Diffusion     & Peptides, Fast Folders              \\
SimpleFold~\citep{wang2025simplefold}       & All-atom       & PDB, ATLAS                & Flow Matching & PDB, ATLAS \\
STAR-MD~\citep{shoghi2026scalable}          & All-atom       & ATLAS                     & Diffusion     & ATLAS, in-house MD \\
Str2Str~\citep{lu2024strstr}                & Backbone       & PDB                       & Diffusion     & BPTI, Fast Folders        \\
TEMPO~\citep{xu2025tempo}                   & Backbone       & ATLAS, mdCATH             & Diffusion     & ATLAS, CATH               \\
TimeWarp~\citep{klein2023timewarp}          & All-atom       & Peptides                  & Normalizing flow & Peptides          \\
UFConf~\citep{fan2024ufconf}                & All-atom       & PDB                       & Diffusion     & Complexes, RAC-47         \\
UniSim~\citep{yu2025unisim}                 & All-atom       & MD, PDB                   & Flow Matching & ATLAS, Small Mol.         \\

% ConfGF~\citep{shi2021confgf}              & Point Cloud    & GEOM                      & Diffusion     & GEOM                      \\
% VonMisesNet~\citep{pmlr-v202-swanson23a}  & SE(3) features & MD                        & Latent LM     & Small Mol.                \\
% Torsional Diffusion~\citep{jing2022torsional} & SE(3) features & GEOM                  & Diffusion     & GEOM                      \\
% GEODIFF~\citep{xu2022geodiff}             & Point Cloud    & GEOM                      & Diffusion     & GEOM                      \\
% GEOMOL~\citep{ganea2021geomol}            & SE(3) features & GEOM                      & OT            & GEOM                      \\
        \bottomrule
    \end{tabularx}
\end{table}

\paragraph{Generative Models for Conformational Ensembles.} A growing body of work directly employs generative models to learn and sample from the equilibrium distribution of protein conformations. Early approaches such as idpGAN~\citep{janson2023idpgan} adopted a transformer-based GAN to generate C-alpha traces, while Str2Str~\citep{lu2024strstr} framed sampling as a translation process inspired by simulated annealing, training solely on static PDB structures via denoising score matching. More recent methods leverage powerful pre-trained structure predictors. AlphaFlow~\citep{jing2024alphaflow} fine-tuned AlphaFold/ESMFold using flow matching objective for conformation ensemble generation. ESMDiff~\citep{lu2025slm} adopts a latent language modeling framework by finetuning ESM3~\citep{hayes2025simulating} with a discrete diffusion objective, producing efficient sequence-conditioned generative models that require no MSAs during inference. Similarly, IDPFold~\citep{zhu2025accurate} combines a protein language model with a diffusion decoder to predict IDP ensembles directly from sequence. Beyond single-state predictors, aSAM~\citep{janson2025asam} employs an all-atom autoencoder with latent diffusion, and its variant aSAMt conditions generation on thermodynamic variables such as temperature. P2DFlow~\citep{jin2025p2dflow} injects a perturbed ESMFold prior into flow matching while conditioning on energy values to sample distinct conformational states. SimpleFold~\citep{wang2025simplefold} takes a minimalist architectural approach, demonstrating that general-purpose transformer blocks trained via flow matching on approximately 9 million distilled structures can achieve competitive folding and ensemble prediction without domain-specific modules such as triangular updates or explicit pair representations. Together, these methods demonstrate the rapid shift from inference-time heuristics to purpose-built generative frameworks that can be trained end-to-end on ensemble data.

% \todo{delete this and disperse the related work to other paragraph.}
\paragraph{Tackling Key Challenges: Physical Priors, Data Scarcity, and PDB Bias.} Despite their promise, purely data-driven generative models often struggle to produce physically realistic ensembles. ConfDiff~\citep{wang2024confdiff} addresses this by incorporating an MD-derived energy prior as a physics-based guide within a force-guided diffusion process, enforcing better adherence to the Boltzmann distribution. Data scarcity is another major obstacle, especially for systems far from experimental coverage. DiG~\citep{zheng2024dig} mitigates this via Physics-Informed Diffusion Pre-training (PIDP), which leverages the Fokker-Planck equation to supervise training directly with energy functions, removing the requirement for steady-state training data. BioEmu~\citep{lewis2025bioemu} further integrates experimental stability measurements to improve thermodynamic property prediction. Meanwhile, UFConf~\citep{fan2024ufconf} confronts the conformational bias of the PDB by reweighting training samples through hierarchical structural clustering, enabling an AlphaFold2-derived diffusion model to explore unseen regions of conformational space. % ScoreMD is discussed in detail in Section 4.2 (ML CG Models).

\paragraph{Extending to Protein-Protein and Protein-Ligand Dynamics.} Generative dynamics modeling is also being extended beyond single-chain proteins to biomolecular interactions. AlphaPPIMD~\citep{wang2024alphappimd} employs a transformer-based latent generative model to sample conformations of protein-protein complexes, trained and evaluated on MD simulations of the barnase-barstar system. DynamicBind~\citep{lu2024dynamicbind} introduces an equivariant diffusion model with a morph-like transformation that iteratively refines AlphaFold-predicted apo structures into ligand-specific holo conformations, effectively performing dynamic docking. PLACER~\citep{ivan2025placer} focuses on ligand dynamics within a protein binding site using a denoising diffusion framework. AnewSampling~\citep{wang2026learning} curated over 15 million protein-ligand conformations and leveraged a quotient-space framework to sample from all-atom equilibrium distributions.

\subsection{Simulation Trajectory generation}\label{subsec:trajectory}

\begin{figure}[!ht]
    \centering
    \includegraphics[width=0.8\linewidth]{figs/fig_learning_from_structure_cropped.pdf}
    \caption{\textbf{Generative modeling of protein structural dynamics from structural data.} \textbf{(A) Conformation ensemble generation.} Generative models learn a distribution over protein structures from structural data. By sampling independent conformations from the learned distribution, the model produces multiple structural realizations that collectively form a conformational ensemble. \textbf{(B) Trajectory generation.} Existing approaches can be broadly categorized into frame transition models that learn long-time molecular dynamics kernels (e.g., MCMC-style transitions), autoregressive dynamics models that predict subsequent conformations sequentially, and one-shot generation models that generate entire trajectories as spatio-temporal sequences.}
    \label{fig:section2}
\end{figure}

\paragraph{Problem Formulation.} Consider a biomolecular system consisting of $N$ atoms. We represent the dynamic evolution of this system as a trajectory of coordinates $\rmX = (\rvx_1, \rvx_2, \dots, \rvx_T) \in \mathbb{R}^{T \times N \times 3}$, where $\rvx_t \in \mathbb{R}^{N \times 3}$ denotes the 3D Cartesian coordinates of all $N$ atoms at time step $t$, and $T$ represents the trajectory length, i.e., the number of frames.  In addition to coordinates, the system is defined by time-invariant atomic features $\rva$, where $\rva \in \mathbb{R}^{N \times d_a}$ represents atom-level attributes including atom type, residue type, molecule-level identifiers that indicate whether an atom belongs to a protein or ligand, etc. Our objective is to learn a generative model $p_\theta(\rmX | \rvx_1, \rva)$ that approximates the true trajectory distribution conditioned on the initial conformation $\rvx_1$ and context $\rva$.

\paragraph{Motivation} In contrast to the conformation generation models discussed in Sec \ref{subsec:ensemble}, which focus primarily on generating conformational ensembles that match equilibrium distributions, trajectory generation methods explicitly model the temporal evolution of protein structures. While conformation generation enables highly efficient parallel sampling of i.i.d. structures, the absence of temporal dependencies precludes the estimation of kinetic observables such as transition rates and folding pathways. Trajectory generation methods address this limitation by emulating the time-ordered dynamics of molecular systems, serving as surrogate models for expensive Molecular Dynamics (MD) simulations.

% \jr{I re-order it a bit to discuss MCMC first, then one-shot, and autoregressive.}
\paragraph{Markov Chain Monte Carlo Methods.} A prominent family of approaches aims to accelerate MD by learning to propose large timestep transitions, effectively acting as surrogate MCMC kernels. These methods typically learn the conditional distribution $p(\rvx_{t+\Delta t} | \rvx_t, \rva)$, where $\Delta t$ is orders of magnitude larger than the femtosecond timestep of MD, enabling rapid exploration of conformational space. ITO~\citep{schreiner2023ito} formulated this problem using conditional denoising diffusion probabilistic models to learn long-time step transitions directly from MD trajectories. TimeWarp~\citep{klein2023timewarp} employs a conditional normalizing flow for the same task, crucially incorporating a Metropolis-Hastings correction step to enforce detailed balance. Subsequent works have explored more sophisticated generative backbones: EquiJump~\citep{costa2024equijump} leverages a two-sided stochastic interpolant that bridges between long-interval timesteps, while DeepJump~\citep{costa2025deepjump} adopts a conditional flow matching framework. F$^3$low~\citep{li2024f} also utilizes conditional flow matching but implements the conditioning mechanism via classifier-free guidance, treating the previous frame as an "initial guess" combined with noise prior through weighted linear interpolation. TEMPO~\citep{xu2025tempo} formulates the transition probability through a one-step SDE that combines deterministic motion with random noise, further incorporating a two-timescale design where a low-resolution model captures slow collective motions that condition a high-resolution model for fine-grained fluctuations. Beyond learning transitions directly in coordinate space, PVB~\citep{yu2026unified} proposes a unified pretraining framework that first maps initial structures to a noised latent space via an encoder-decoder, then transports them toward stage-specific targets using augmented bridge matching. This design enables consistent training across both single-structure pretraining and paired-trajectory supervision for step transition finetuning. GLDP~\citep{sengar2026beyond} similarly operates in a compressed latent representation of all-atom proteins, learning dynamics in this reduced space before decoding back to atomistic trajectories; the authors benchmark three transition operators within this latent space: score-guided Langevin dynamics, a linear Koopman operator, and a standard MLP.

A distinct line of MCMC methods focuses specifically on transitions between metastable states rather than frame-to-frame dynamics. ScooBDoob~\citep{zhang2025scoobdoob} formulates a discrete bridge-matching framework that tilts Markov state model transition rates via Doob's $h$-transform, generating optimal stochastic paths between prescribed initial and terminal ensembles. Mars-FM~\citep{kapusniak2025mars} similarly samples transitions between metastable states defined by an MSM, employing a flow matching framework to generate these state-to-state trajectories without learning every intermediate frame, offering a complementary approach to traditional MD emulation.

\paragraph{One-Shot Trajectory Generation.} Rather than iteratively stepping through time, an alternative paradigm generates complete trajectories in a single forward pass. Pioneered by MDGen~\citep{jing2024mdgen}, this approach frames molecular dynamics trajectories as "molecular videos" and learns a generative model that produces the full time series of 3D structures—an approach we term \textit{one-shot trajectory generation}. These models directly generate the entire sequence $\rmX = (\rvx_1, \rvx_2, \dots, \rvx_T)$ conditioned on the initial conformation $\rvx_1$ and atomic context $\rva$. This paradigm enables a unified architecture capable of performing multiple tasks beyond forward simulation—such as transition path sampling—through different masking strategies applied during training and inference. Building on this foundation, BioMD~\citep{feng2026biomd} extends the video analogy to protein–ligand systems, while PTraj-Diff~\citep{xu2025ptraj-diff} adapts the framework to a diffusion architecture with tensor product attention for improved efficiency and a BERT~\citep{devlin2018bert} encoder for enhanced temporal modeling, with a focus on protein–protein complexes. DynaFold~\citep{fan2025dynafold} further advances the trajectory-as-video concept by operating in a compressed latent space, achieving greater computational and data efficiency through one-shot generation in the latent trajectory space. AlphaFolding~\citep{cheng2025alphafolding} treats the trajectory as a 4D structure and employs a 4D diffusion model conditioned on the initial 3D conformation, directly generating time-ordered coordinate sequences.

\paragraph{Autoregressive Methods.} A complementary line of work adopts autoregressive formulations for trajectory generation, where each subsequent frame is predicted conditioned on previously generated ones. ConfRover~\citep{shen2025confrover} encodes historical frames into a latent representation sequence and predicts a subsequent latent state, from which the next conformation is decoded by a diffusion-based decoder conditioned on this latent. STAR-MD~\citep{shoghi2026scalable} addresses the computational bottlenecks of such architectures by introducing an efficient joint spatio-temporal attention mechanism, enabling scalable autoregressive dynamics modeling. ATMOS~\citep{shi2026atmos} adapts State Space Models to this task, capturing long-range temporal dependencies with linear complexity in trajectory length and extending the framework to protein–ligand systems. MD-LLM-1~\citep{murtada2025mdllm1} takes a radically different path by leveraging large language models: it tokenizes protein structures and fine-tunes a modern LLM (Mistral 7B) via LoRA for trajectory generation. BioKinema~\citep{feng2026physically} performs all-atom biomolecular dynamics generation using forecasting-interpolation similar to BioMD~\citep{feng2026biomd}, yet also supports the autoregressive paradigm with the Spatial-Temporal attention. UniSim~\citep{yu2025unisim} proposes a unified cross-domain simulator that first learns atomic representations from diverse molecular data via multi-head pretraining, then employs a stochastic interpolant framework with a force guidance module for rapid adaptation, demonstrating competitive performance across small molecules, peptides, and proteins.

\section{Learning from Energy Signals}\label{sec:energy_signals}

% method name; variant (limitation of BG);  training/test system (peptides, BPTI, fast folders, etc.); transferrability; equivariance property; likelihood 

\begin{table}[htbp]
    \centering
    \scriptsize
    \caption{Overview of Boltzmann generators and related energy-driven samplers. For each method, we report the molecular representation (\textbf{Repr.}), learning objective or model family (\textbf{Obj.}), demonstrated system scale (\textbf{Scaling}), and the force field or energy model used in evaluation. Abbreviations: AA = all-atom; Cart. = Cartesian; int. = internal coordinates; mixed = mixed-coordinate representation; NF = normalizing flow; FM = flow matching.}
    \label{tab:lit_review_energy_signals}
    \begin{tabularx}{\linewidth}{@{}l l l l X@{}}
        \toprule
        \textbf{Name} & \textbf{Repr.} & \textbf{Obj.} & \textbf{Scaling} & \textbf{Force Field} \\
        \midrule
        BG~\citep{noe2019bg}
        & AA mixed
        & NF
        & BPTI (58 res.)
        & Amber99SB-ILDN\\
        
        TBG~\citep{klein2024tbg}
        & AA Cart.
        & FM
        & Dipeptides
        & GFN2-xTB, Amber99SB-ILDN, Amber14\\ 

        SBG~\citep{tan2025sbg}
        & AA Cart.
        & NF 
        & Up to 10 res.
        & Amber99SB-ILDN, Amber14\\
        
        PROSE~\citep{tan2025prose}
        & AA Cart.
        & NF
        & Up to 8 res.
        & Amber14\\
        
        Eq.\ FM~\citep{klein2023equivariant}
        & AA Cart.
        & FM
        & Dipeptides, LJ-55
        & GFN2-xTB, LJ, Amber99SB-ILDN\\
        
        Smooth NF~\citep{kohler2021smooth}
        & AA int.
        & NF
        & Dipeptides
        & Amber99SB-ILDN\\

        FAB~\citep{midgley2023flow}
        & AA int.
        & NF
        & Dipeptides
        & Amber96\\

        Scalable BG~\citep{kim2024scalable}
        & AA mixed
        & NF
        & Up to 56 res.
        & Amber14\\

        iDEM~\citep{idem}
        & AA Cart.
        & Diffusion
        & LJ-55
        & LJ\\

        BNEM~\citep{ouyang2025bnemboltzmannsamplerbased}
        & AA Cart.
        & Diffusion
        & LJ-55
        & LJ\\

        PITA~\citep{akhound-sadegh2025progressive}
        & AA Cart.
        & Diffusion
        & Tripeptides
        & LJ, Amber14\\

        TA-BG~\citep{schopmans2025temperatureannealed}
        & AA int.
        & NF
        & Hexapeptides
        & Amber96, Amber99SB-ILDN\\

        EWFM (i/a)~\citep{dern2025energyweightedflowmatchingunlocking}
        & AA Cart.
        & FM
        & LJ-55
        & LJ\\

        CMT~\citep{klitzing2026learning}
        & AA int.
        & NF
        & Hexapeptides
        & Amber99SB-ILDN, Amber96\\

        RegFlow/FORT~\citep{rehman2025efficient}
        & AA Cart.
        & NF
        & Tetrapeptides
        & Pre-trained CNF/OT\\

        LDR~\citep{schopmans2026ldr}
        & AA mixed
        & Any
        & Hexapeptides
        & Amber96, Amber99SB-ILDN\\
        
        % (i,a)-EWFM & AA Cartesian & None & FM & Up to LJ-55 &  & None \\
        
        % - & - & - & - & - & - & - \\
        % SBG & AA Cartesian & MD & NF + Inference-time SMC & Up to hexapeptide &  & 100-1000 \\
        % BG & AA Internal & None & NF & Up to 58 residues (BPTI) & & Importance Reweighting \\
        % TBG & AA Cartesian & MD & FM + EGNN & Up to dipeptides & & None \\
        % PROSE & AA Internal & ManyPeptidesMD & NF + TarFlow & Up to octapeptides \& Chignolin &  & Importance Reweighting \\
        % Eq. FM & AA Cartesian & High-temp. MD  & OT-FM + $E(3)$-equiv. GNN & Up to dipeptide, LJ-55 &  & Importance Reweighting \\
        % Smooth NF & AA Internal & MD+Forces & NF & Up to dipeptide &  & None \\
        % FAB & AA Cartesian & None & AIS + NF & Up to dipeptide &  & 1-10 \\
        % Scalable BG & AA Internal & Equilibrium MD & NF & Up to 56 residues (villin HP35, Protein G) &  & 0-10 \\
        % iDEM & AA Cartesian & None & Diffusion + $SE(3)$-equiv. GNN & Up to LJ-55 &  & None \\
        % BNEM & AA Cartesian & None & Diffusion + Energy & Up to LJ-55 & LJ-55, Double-well & 100-1000 \\
        % TA-BG & AA Internal & None & NF & Up to hexapeptide &  & 10-100 \\
        % PITA & AA Cartesian & High-temp. MD & Diffusion w/ Temp. + Prog. Distillation & Up to tripeptide &  & 10-100 \\
        % TorsionNet & CG Internal & None & PPO + GNN w. LSTM memory & Up to 150-200 atoms &  & 10-1000 \\

        \bottomrule
    \end{tabularx}
\end{table}

% \todo{@yashi + @jiarui}
% This section lists the methods including (but not limited to): Boltzmann Generators (\eg, BG, TBG); amortized ML samplers (equilibrium distribution sampling, \eg, GFlowNet); Incorporate physical supervision into the model (finetuning).

In this section, we provide a definition of the Boltzmann generator and present how different methods attempt to produce models that are able to sample and estimate observables of interest. To put simply, a \emph{Boltzmann generator} is a conformation generative model that also has its own likelihood model $\tilde{p}(\mathbf{x})\approx p(\mathbf{x})$. Estimating observables of interest is then possible by importance sampling with respect to the true distribution $p(\mathbf{x})$.  

\begin{figure}[!ht]
    \centering
    
    % --- Top subfigure ---
    \begin{subfigure}{1.0\linewidth}
        \centering
        \includegraphics[width=0.9\linewidth]{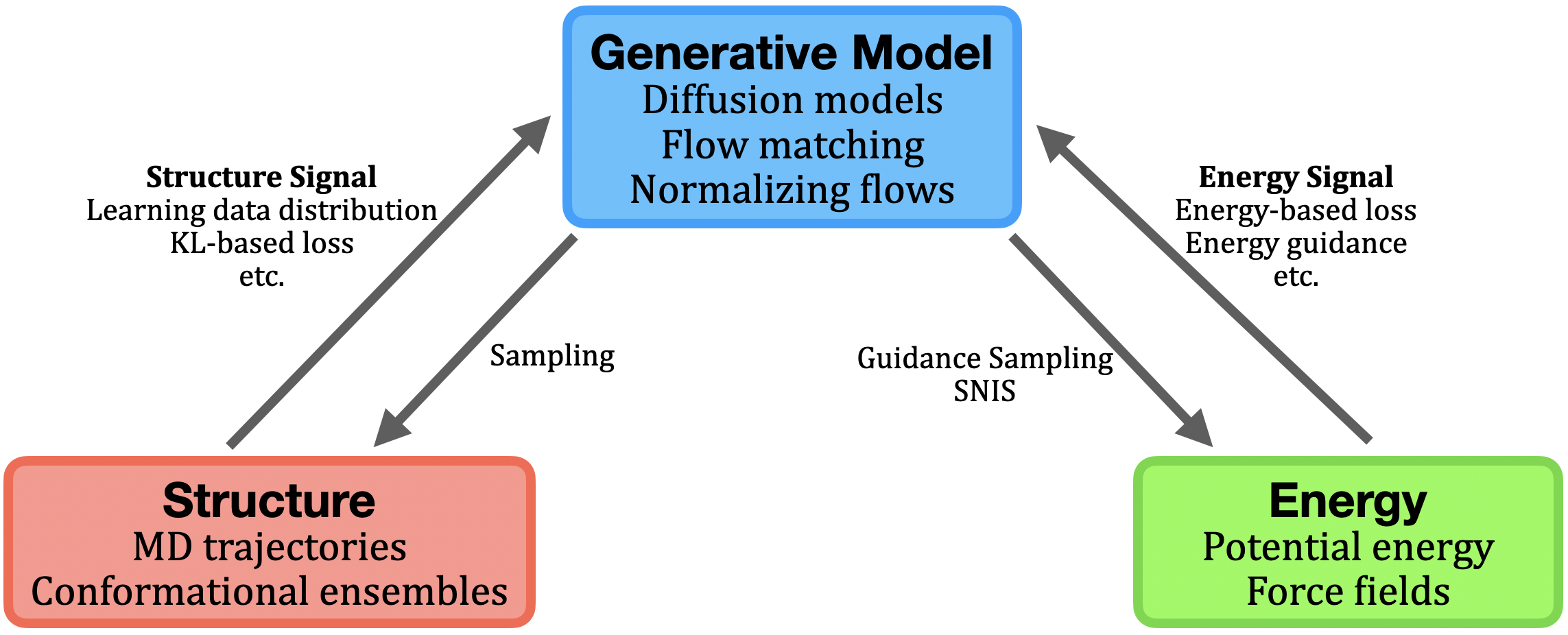}
        \caption{Structure--Energy--Model Interactions}
        \label{fig:sec3:process}
    \end{subfigure}
    
    \vspace{0.5em}
    
    % --- Bottom subfigure ---
    \begin{subfigure}{1.0\linewidth}
        \centering
        \includegraphics[width=1.0\linewidth]{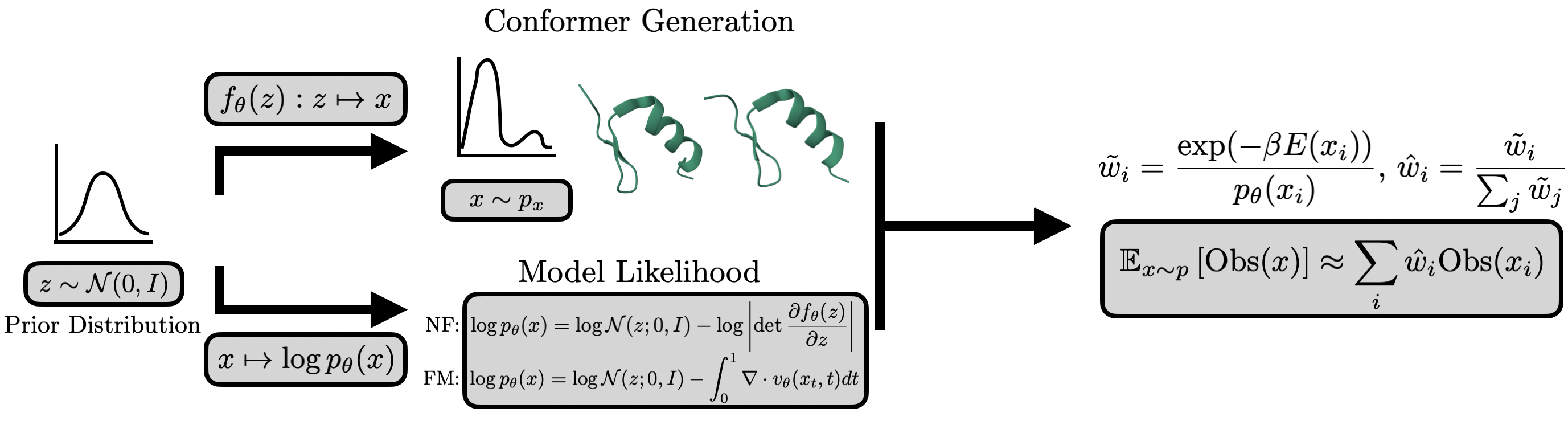}
        \caption{Boltzmann Generator Sampling}
        \label{fig:sec3:bg}
    \end{subfigure}
    
    \caption{\textbf{Overview of data--energy--model interactions and Boltzmann generator sampling.}
    \textbf{Top}: Structural information from MD trajectories and conformational ensembles provides a structure signal for training generative models, which then sample new structures. Potential energies and force fields can provide a training or inference signal through energy-based losses or guidance, which can also be incorporated during generation via guidance sampling or self-normalized importance sampling (SNIS).
    \textbf{Bottom}: A latent variable $z \sim \mathcal{N}(0,I)$ is transformed by a learned generator $f_\theta: z \mapsto x$ to produce molecular conformers $x$ drawn from a tractable model distribution ($p_\theta(x)$). The model likelihood can be evaluated either with the change-of-variables formula for a normalizing flow or, for continuous flows, by integrating the divergence of the velocity field. Generated samples are then importance reweighted according to ($\tilde w_i = \exp[-\beta E(x_i)]/p_\theta(x_i)$) and used to estimate Boltzmann ensemble averages ($\mathbb{E}_{x\sim p}[\mathrm{Obs}(x)] \approx \sum_i \hat w_i \mathrm{Obs}(x_i)$).
}
    
    \label{fig:sec3:combined}
\end{figure}

% \begin{figure}[!ht]
%     \centering
%     \includegraphics[width=0.95\linewidth]{figs/sec3_process_diagram.png}
%     \caption{\textbf{Placeholder Figure: }Process Diagram for Data–Energy–Model Interaction in Generative Protein Dynamics Modeling}
%     \label{fig:sec3:process}
% \end{figure}

\paragraph{Problem Formulation.}
In many protein dynamics settings, we are given an energy function $E(\rvx)$ (e.g., a classical force field, an implicit-solvent model, or a learned potential) that defines a target equilibrium distribution over conformations,
\begin{equation*}
    p(\rvx) \;=\; \frac{1}{Z}\exp\!\big(-\beta E(\rvx)\big), 
    \qquad Z=\int \exp\!\big(-\beta E(\rvx)\big)\, d\rvx,
\end{equation*}
where $\beta = (k_BT)^{-1}$. The goal is to \emph{efficiently sample} $\rvx \sim p(\rvx)$ and to \emph{estimate observables} $\mathbb{E}_{p}[f(\rvx)]$ and free-energy differences, ideally for protein systems whose energy landscapes are high-dimensional and strongly multimodal.

\paragraph{Why learn from energy signals?}
Molecular dynamics (MD) and MCMC methods provide asymptotically exact sampling but often require prohibitively long trajectories to cross metastable barriers. In contrast, energy evaluations are ubiquitous and physically grounded: for any proposed conformation $\rvx$, we can typically compute $E(\rvx)$ and often its gradient $\nabla_{\rvx}E(\rvx)$ (forces).
This enables a complementary paradigm to purely data-driven ensemble learning: \emph{train generative samplers using energy/force supervision}, either (i) entirely without equilibrium data (simulation-free inner loops), or (ii) as a post-training alignment signal that corrects dataset bias and improves thermodynamic faithfulness.
A central practical consideration is that energy and force evaluations can be expensive, so methods are often compared by the number of energy evaluations per (effectively independent) sample.

\paragraph{Evaluation and correction with self-normalized importance sampling (SNIS).}
A useful separation is between \emph{proposal generation} and \emph{thermodynamic correctness}. When a learned sampler provides a tractable (or accurately estimated) proposal density $\tilde{p}_\theta(\rvx)$, samples $\{\rvx_i\}_{i=1}^N \sim \tilde{p}_\theta$ can be reweighted to estimate expectations under the Boltzmann distribution without knowing the partition function $Z$. Define
\begin{equation*}
    \tilde{w}_i \;=\; \frac{\exp\!\big(-\beta E(\rvx_i)\big)}{\tilde{p}_\theta(\rvx_i)},
    \qquad
    \hat{w}_i \;=\; \frac{\tilde{w}_i}{\sum_{j=1}^N \tilde{w}_j},
\end{equation*}
so that
\begin{equation*}
    \mathbb{E}_{p}[f(\rvx)]
    \;\approx\;
    \sum_{i=1}^N \hat{w}_i\, f(\rvx_i).
\end{equation*}
By choosing $f$ appropriately, SNIS can estimate mean potential energy, metastable-state populations and free-energy differences, partition-function ratios, heat capacities from energy fluctuations, and free-energy profiles along reaction coordinates. It also supports \emph{energy-distribution metrics}: for energies $u_i = E(\rvx_i)$ and an energy bin $B$,
\begin{equation*}
    \widehat{p}(u \in B)
    \;=\;
    \sum_{i:\, u_i \in B} \hat{w}_i,
    \qquad
    \widehat{F}(u \in B)
    \;=\;
    -\beta^{-1}\log \widehat{p}(u \in B) + C,
\end{equation*}
where $C$ is an arbitrary additive constant. In practice, weighted energy histograms or free-energy curves can be compared against reference simulations, along with discrepancies in low-order moments or 1D marginals.

A key diagnostic for the reliability of this reweighting is the \emph{effective sample size}
\begin{equation*}
    \mathrm{ESS}
    \;=\;
    \frac{1}{\sum_{i=1}^N \hat{w}_i^2}
    \;=\;
    \frac{\big(\sum_{i=1}^N \tilde{w}_i\big)^2}{\sum_{i=1}^N \tilde{w}_i^2},
\end{equation*}
often reported per energy/force evaluation. Low ESS indicates poor overlap between $\tilde{p}$ and $p$, leading to high-variance estimates; in such cases, annealing- or SMC/AIS-style corrections provide alternative routes to unbiasedness, at the cost of additional energy evaluations.

\paragraph{Scope of this section.}
We survey methods that explicitly leverage energy signals for equilibrium sampling and (in some cases) dynamics:
\emph{(i) Boltzmann generators and related likelihood models} that learn expressive proposals with (approximate or exact) likelihoods and correct residual bias via reweighting or annealing;
\emph{(ii) physics-aware adaptation}, which incorporates physical energy signals during post-training or at inference time. The former fine-tunes pretrained protein generators using energy-based objectives while the latter uses energies to guide sampling. Key design axes include symmetry handling, the bias--variance tradeoff of reweighting, and scalability. 

% \iffalse
% \subsection{Amortized inference in ML}
% TBA. Put GFlownet related work here: Learning from unnormalized density / energy functions.
% \begin{itemize}
%     \item GFN for Conformation~\citep{VOLOKHOVA20241038}
%     A general outline/survey paper that describes the methodology of GFlowNets and how they have been or could be used to sample molecular conformations. So far, there does not seem to be any GFN methods on larger scale molecular systems at the peptide scale. 
%     \item BoltzNCE~\citep{aggarwal2025boltznce}
%     BoltzNCE trains an auxiliary EBM such that the score (gradient with respect to inputs, not model parameters) satisfies the stochastic interpolant equation. Notably, the framework uses two networks, a primary diffusion network (Boltzmann Emulator) for sampling conformations, and a secondary EBM (likelihood estimator) to compute the likelihood of the conformation in one forward pass. These two networks do not see each other during training. 
% \end{itemize}
% \fi

\subsection{Boltzmann Generators}\label{subsec:boltzmann}

% \begin{figure}[!ht]
%     \centering
%     \includegraphics[width=1.0\linewidth]{figs/sec3_boltzmann_generator.png}
%     \caption{\textbf{Placeholder Figure: }Training Diagram for Boltzmann Generator}
%     \label{fig:sec3:bg}
% \end{figure}

\paragraph{Normalizing flows as likelihood-based generators.}
Normalizing flows (NFs) are generative models that construct a flexible density on configurations by transforming a simple base distribution through an \emph{invertible} map.
Concretely, let $\rvz \sim \calN(0,I)$ and define
\begin{equation}
    \rvx = f_\theta(\rvz), \qquad \rvz = f_\theta^{-1}(\rvx),
\end{equation}
where $f_\theta:\mathbb{R}^d\!\to\!\mathbb{R}^d$ is bijective and differentiable.
The model density is then available via the change-of-variables formula,
\begin{equation}
    \tilde{p}_\theta(\rvx)
    \;=\;
    \calN\!\big(f_\theta^{-1}(\rvx);0,I\big)\;
    \left|\det \frac{\partial f_\theta^{-1}(\rvx)}{\partial \rvx}\right|
    \;=\;
    \calN(\rvz;0,I)\;
    \left|\det \frac{\partial f_\theta(\rvz)}{\partial \rvz}\right|^{-1}.
\end{equation}
This combination of \emph{fast sampling} ($\rvz \mapsto \rvx$) and a \emph{tractable likelihood} ($\log \tilde{p}_\theta(\rvx)$) enables Boltzmann-generator-style
reweighting/annealing corrections, e.g.\ $w(\rvx)\propto \exp(-\beta E(\rvx))/\tilde{p}_\theta(\rvx)$.

\paragraph{Architectures and likelihood cost.}
Discrete-time flows (e.g., coupling-layer models such as RealNVP) are built so that the Jacobian determinant is cheap to compute, often triangular by design.
Continuous-time flows (CNFs) instead define an ODE $\tfrac{d\rvx_t}{dt}=v_\theta(\rvx_t,t)$; their likelihood requires integrating a divergence term
$\tfrac{d}{dt}\log \tilde{p}_t(\rvx_t) = -\nabla_{\rvx}\!\cdot v_\theta(\rvx_t,t)$, which is inaccurate and computationally intractable in general. 
More recent flow-matching objectives sidestep explicit likelihood training by learning $v_\theta$ to match a prescribed transport, trading exact likelihood access for improved scalability.
For molecular systems, models need coordinate choices (Cartesian vs.\ internal) and respective symmetry handling ($SE(3),\ E(3)$), which impact sample quality and data efficiency non-trivially.

\paragraph{Classical BGs and exact-likelihood descendants.}
Fast direct samples should come with an explicit proposal density that supports principled correction. The original BG framework used RealNVP-style~\citep{dinh2016density} invertible flows so that generated samples could be corrected by SNIS or annealing-based reweighting~\citep{noe2019bg}. After the introduction of TarFlow~\citep{zhai2025normalizing}, a transformer-based scalable flow network, SBG~\citep{tan2025sbg} utilized it to emphasize inference-time correction, coupling the learned proposal with annealed importance sampling to improve overlap on larger peptide systems. PROSE~\citep{tan2025prose} extends the same exact-likelihood recipe to transferable multi-system training, conditioning on atom-level and sequence-level features so that zero-shot equilibrium sampling and temperature transfer remain compatible with SNIS reweighting.

\paragraph{Diffusion-based amortized samplers.}
Diffusion-based Boltzmann samplers trade tractable likelihoods for flexible, symmetry-aware generators trained directly from energetic supervision. iDEM learns a simulation-free, $SE(3)$-equivariant diffusion model from energies and gradients through denoising energy matching and a replay-buffer outer loop that refreshes informative low-energy conformations~\citep{idem}. BNEM preserves the simulation-free inner-loop philosophy but parameterizes time-dependent noised energies rather than scores, reducing target variance at the cost of an extra backward pass~\citep{ouyang2025bnemboltzmannsamplerbased}. PITA combines diffusion, temperature conditioning, and progressive distillation, using high-temperature MCMC data and Feynman--Kac-based annealing to bootstrap progressively lower-temperature samplers \citep{akhound-sadegh2025progressive}.

\paragraph{Energy-guided likelihood-free generative models and stable transport learning.}
A broad middle ground relaxes exact-likelihood training in favor of more expressive, energy-aware, and scalable transports. TBG replaces MLE-style BG training with EGNN-based, roto-permutation-equivariant flow matching in all-atom Cartesian coordinates, often operating without model likelihood due to high computational costs~\citep{klein2024tbg}. Equivariant Flow Matching pursues a closely related $E(3)$-equivariant transport formulation that prioritizes geometric inductive bias and scalable flow matching over exact density evaluation~\citep{klein2023equivariant}. Smooth Normalizing Flows address a complementary concern by constructing infinitely differentiable, compactly supported transformations that behave better for mixed Euclidean and periodic coordinates when stable derivatives or forces matter~\citep{kohler2021smooth}. FAB shows that useful Boltzmann proposals can be learned from energies alone by coupling a normalizing flow with annealed importance sampling and optimizing a loss tied to importance-weight variance~\citep{midgley2023flow}. TA-BG and EWFM tackle the same overlap problem through annealed, energy-aware objectives, with TA-BG iteratively cooling a high-temperature flow by importance reweighting and forward-KL training and EWFM introducing amortized and annealed energy-reweighted flow-matching losses for rougher landscapes~\citep{schopmans2025temperatureannealed,dern2025energyweightedflowmatchingunlocking}. Scalable BGs push BG-style learning to substantially larger proteins through long-context-inspired architectures and a staged transition from maximum likelihood to a 2-Wasserstein objective on backbone distance matrices~\citep{kim2024scalable}. CMT, RegFlow/FORT, and LDR address training pathologies from different angles by constraining intermediate transport to reduce mass teleportation, replacing fragile likelihood objectives with regression and self-consistency regularization, and adding off-policy log-dispersion with explicit energy labels to improve data efficiency on fixed datasets~\citep{klitzing2026learning,rehman2025efficient,schopmans2026ldr}.

%\paragraph{Related directions beyond exact-likelihood BGs.}
%Some neighboring methods do not fit the strict Boltzmann-generator definition, but are still relevant because they learn sampling policies from energy feedback. TorsionNet~\citep{torsionnet-rl}, for example, formulates conformer search as a sequential reinforcement learning problem in coarse-grained internal coordinates, where a GNN with LSTM memory proposes torsion-angle updates and receives energy-based rewards after MMFF94 relaxation. Its non-stationary reward discourages duplicate conformers, and optimization against the Gibbs score explicitly connects RL objectives to ensemble quality. E-NEQ~\citep{holdijk2026learning} is similarly adjacent to the BG literature but focuses on learning escorted switching protocols for relative free-energy estimation rather than direct one-shot equilibrium sampling. By combining conditional flow matching, conditional density matching, and multistate flow graphs, it learns transitions that reduce dissipation and improve free-energy calculations.

\subsection{Physics-aware adaptation of generative models}\label{subsec:physics_adapt}

\paragraph{Motivation.}
% The recent wave of protein diffusion and flow models is often trained on structural databases or MD trajectories whose empirical distribution does not exactly match a target force field, temperature, or experimental thermodynamic ensemble. This creates an important opportunity for \emph{post hoc} physical alignment: rather than training a Boltzmann sampler from scratch, one can begin with a strong pretrained generator and use energies, forces, constraints, or collective variables to steer its distribution toward physically meaningful ensembles. In contrast to classical BGs, many of these methods do not require explicit likelihoods and instead rely on ranking, weighting, guidance, or differentiable projection.
Recent protein diffusion and flow models are trained on data distributions that may not match the target thermodynamic ensemble, motivating \emph{post hoc} physical alignment: starting from a pretrained generator, energies, forces, or constraints can be used to steer samples toward physically meaningful distributions, often without requiring explicit likelihoods. For a base model generating samples from $\tilde{p}_\theta(x)$, we wish to sample from the \emph{tilted density} 
$$p_{\text{adapt}}(x)\propto\tilde{p}_\theta(x)\exp(-\lambda r(x))$$ 
for some guidance strength $\lambda>0$ and reward function $r(x)$. This reward function is often chosen to be a surrogate of the energy, validity, or other desirable thermodynamic properties. 

\paragraph{Post-training alignment.}
Recent work treats strong pretrained protein generators as starting points and then injects physical information \emph{post hoc}. EPO~\citep{sun2025epo} and EBA~\citep{lu2025eba} both use physics-based energies to bias generators toward lower-free-energy ensembles without explicit likelihoods: EPO uses list-wise preference optimization with a tractable upper bound on trajectory likelihood ratios, whereas EBA uses a mini-batch Boltzmann-factor objective that approximately preserves energy-implied probability ratios and an energy-weighted list-wise denoising loss. 

\paragraph{Inference-time correction and steering.}
These methods use physical feedback to improve realism, enforce constraints, steer exploration, and extend pretrained generators towards enhanced-sampling or coarse dynamical regimes. 
ConfDiff~\citep{wang2024confdiff} adds force guidance from an MD prior, Gauss--Seidel Projection~\citep{chen2026physically} inserts an implicitly differentiated projection step that enforces steric and geometric validity and can reduce denoising steps, Metadiffusion~\citep{lam2026metadiffusion} biases pretrained generators along chosen collective variables or experimental observables, and WT-ASBS~\citep{nam2026enhancing} adds a sequential collective-variable bias that promotes rare-event exploration while permitting reweighting.

\paragraph{Section summary.}
Across this section, the field can be read as a spectrum. At one end, exact-likelihood BGs provide the strongest route to unbiased thermodynamic estimation; in the middle, likelihood-free generative models use energy supervision to amortize sampling while relaxing likelihood requirements; at the other end, physics-aware adaptation retrofits thermodynamic information onto pretrained structural generators. The central open problem is to combine the \emph{correction guarantees} of classical BGs, the \emph{symmetry-aware scalability} of modern diffusion and flow models, and the \emph{transferability} of foundation-model-style pretraining within a single framework.

\section{Learning for MD simulations}\label{sec:learn_mds}

% \todo{@haocheng + @xixian}. This section lists: Machine Learning potentials (force fields) and learnable collective variables.

\begin{figure}[!ht]
    \centering
    \includegraphics[width=1.0\linewidth]{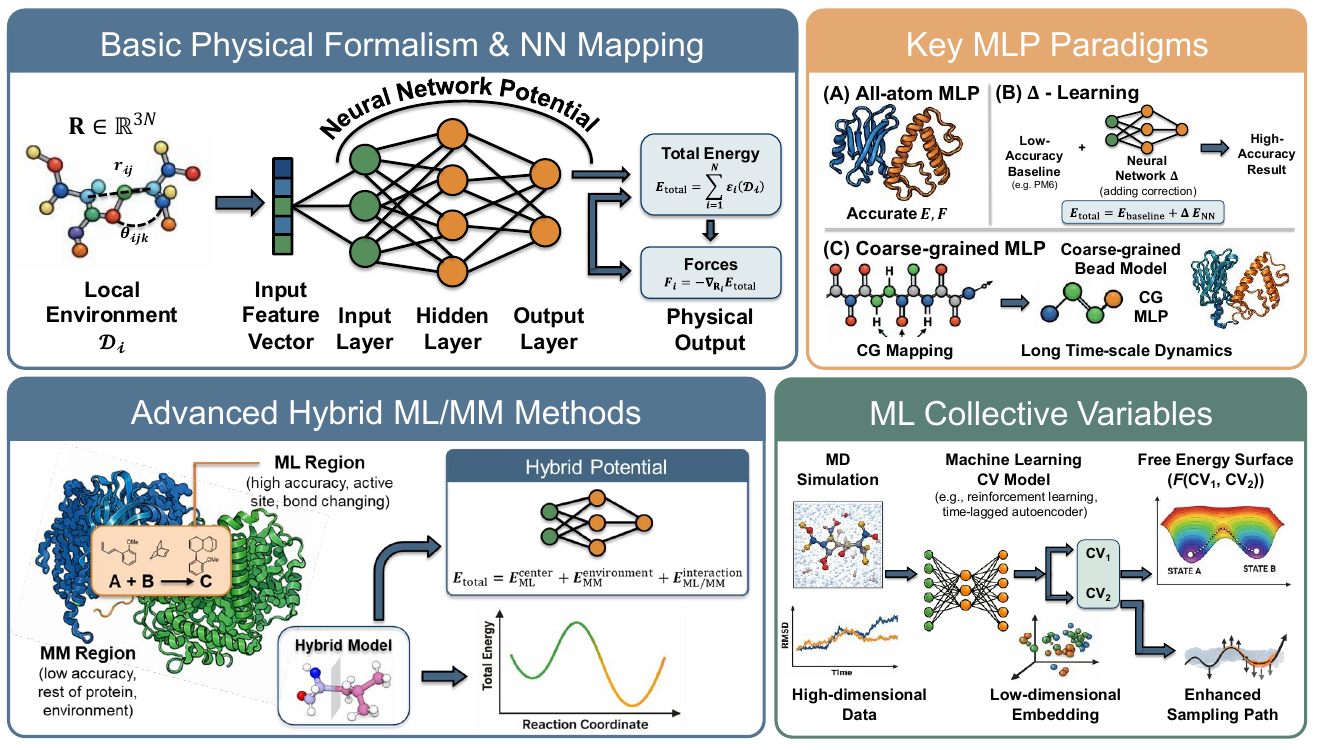}
    \caption{Machine Learning Potentials (MLPs) and Collective Variables (CVs) for Molecular Dynamics. \textbf{Top Left:} Mapping of local atomic environments ($r_{ij}, \theta_{jik}$) through neural networks to predict total energy $E_{MLP}$ and forces $F_i$. \textbf{Top Right:} Core paradigms including Full MLP, $\Delta$-learning for accuracy correction, and coarse-grained models for extended timescales. \textbf{Bottom Left:} Hybrid ML/MM schemes for enzymatic reactions, coupling high-precision ML active sites with MM environments.\textbf{ Bottom Right:} ML-based dimensionality reduction (e.g., Autoencoders) to identify collective variables and reconstruct Free Energy Surfaces.}
    \label{fig:section4}
\end{figure}

This section provides an overview of machine learning methods for biomolecular dynamics simulation. Current methods mainly focus on basic physical formalism and neural network mapping, enabling the transformation from local atomic environments (e.g., distances $r_{ij}$ and angles $\theta_{jik}$) to individual atomic energy contributions $E_i$ and accelerating the exploration of potential energy surface (PES) of complex systems.

% \jr{\textbf{Note for MLP}:Draw a fig, contain MLP/CGMLP/Enzyme/CV}

\begin{table}[htbp]
    \centering
    \scriptsize
    \caption{Overview of related methods of \textbf{machine learning potentials} including system scale.}
    \label{tab:lit_review_extended}
    \begin{tabularx}{\linewidth}{@{}l l X X l@{}}
        \toprule
        \textbf{Name} & \textbf{Rep.} & \textbf{Training Data} & \textbf{Test Systems} & \textbf{System Size} \\
        \midrule
        Espaloma~\citep{takaba2024espaloma} & All-Atom & SPICE, peptides & Small molecules, proteins, nucleic acids & Scalable \\
        AI$^2$BMD~\citep{wang2024ai2bmd} & All-Atom & Protein units & Proteins & $>$10,000 atoms \\
        GEMS~\citep{unke2024gems} & All-Atom & GEMS & Proteins & $>$25,000 atoms \\
        AQuaRef~\citep{zubatyuk2025aquaref} & All-Atom & Polypeptides, complexes & Proteins & Full Protein \\
        SO3LR~\citep{kabylda2025so3lr} & All-Atom & GEMS, QM7-X, SPICE & Complex biosystems & $>$200,000 atoms \\
        CGnets~\citep{wang2019cgnet} & CG & System-specific & Specific proteins & Small proteins \\
        CGSchNet~\citep{charron2025cgschnet} & CG & mdCATH, peptide & Proteins & Scalable \\
        PCCG~\citep{ding2022pccg} & CG & System-specific & Protein domains & Protein domains \\
        TorchMD-Net~\citep{majewski2023torchnet} & AA/CG & QM9, MD17, Protein folding & Proteins & Scalable \\
        Flow-Matching~\citep{Kohler2023flowmatching} & CG & System-specific & Proteins & Small to Medium \\
        InvertibleCG~\citep{chennakesavalu2023thermodynamicconsistency} & CG & System-specific & Specific proteins & Scalable \\
        2-for-1-diffusion~\citep{arts20232for1diffusion} & CG & System-specific & Proteins & Scalable \\
        ScoreMD~\citep{plainer2025consistent} & CG & Fast-folding, dipeptides & Proteins & Scalable \\
        GNN-LRP~\citep{bonneau2025gnnlrp} & Analysis & System-specific & GNN potential & N/A \\
        \bottomrule
    \end{tabularx}
\end{table}

% \subsection{Unclassified}
% \begin{itemize}
%     \item SMA-MD~\citep{VigueraDiez_2024} SMA-MD(surrogate model-assisted molecular dynamics) is a practical strategy for equilibrium conformational sampling that combines a torsional generative surrogate (to accelerate mixing of slow degrees of freedom) with Boltzmann reweighting and short parallel MD refinement (to recover unbiased local fluctuations).   Their main contributions are the formulation of this two-stage workflow, systematic benchmarking against MD and replica exchange, the release of the MDQM9-nc dataset (trajectories for 12,530 non-cyclic QM9 molecules), and a thermodynamics case study on solvation free energies.   Historically, SMA-MD is notable less for proposing a new potential and more for codifying a general sampling paradigm: using fast generative proposals to cross torsional barriers while preserving statistical correctness through reweighting and physics-based refinement, offering a scalable route to production-quality equilibrium ensembles when brute-force MD is limited by metastability. 
% \end{itemize}

\subsection{Machine Learning Potentials}\label{subsec:mlp}

\paragraph{Problem Formulation.}
The fundamental challenge in molecular modeling is to determine the potential energy surface (PES) $E(\mathbf{R})$ and its resulting forces $\mathbf{F} = -\nabla E(\mathbf{R})$ for a system of $N$ atoms with coordinates $\mathbf{R} \in \mathbb{R}^{3N}$. In the context of protein dynamics, this PES dictates the Boltzmann distribution $p(\mathbf{R}) \propto \exp(-E(\mathbf{R})/k_B T)$. The goal of machine learning potentials (MLPs) is to find a functional mapping $f_\theta: \{\mathbf{R}_i, Z_i\}_{i=1}^N \to E$ that approximates the high-level quantum mechanical (QM) energy while remaining efficient enough for long-time sampling and stable molecular dynamics (MD), where $Z_i$ stands for atom types.

\paragraph{Motivation.}
Historically, the trade-off between accuracy and efficiency has defined the limits of molecular simulation. At one extreme, \textit{ab initio} QM methods provide high fidelity by solving the Schrödinger equation but are computationally restricted to systems of a few hundred atoms. At the other, empirical force fields (FFs) simplify interactions into fixed functional forms (e.g., harmonic bonds, Lennard-Jones potentials). While FFs enable microsecond-scale simulations, their rigid parameters often fail to capture complex chemical phenomena such as bond breaking, polarization, or many-body effects. Hybrid QM/MM (Quantum Mechanics/Molecular Mechanics) methods attempt to bridge this gap by treating the active site with QM and the environment with MM; however, they still suffer from high computational cost.

The emergence of MLPs offers another choice. By leveraging deep neural networks as flexible interpolators, MLPs can learn the underlying physics of QM data with near-QM accuracy while operating at speeds approaching empirical force fields. Specifically, in complex environments like protein-solvent systems, MLPs can explicitly model non-bonded interactions and many-body expansions that are typically neglected or oversimplified in traditional MM. This technique enables the exploration of vast conformational landscapes with chemical accuracy, overcoming the previous limitations of both purely empirical and purely quantum approaches.

\paragraph{Methods.}
The fundamental objective of MLPs is to construct an efficient mapping from the atomic configuration, defined by atomic positions $\{R_i\}$ and atomic types $\{Z_i\}$, to the PES. Unlike traditional force fields that rely on fixed functional forms, modern MLPs leverage deep neural networks, usually GNN and MLP, to learn the complex many-body interactions directly from QM data.

Most MLPs for biomolecular systems operate on the principle of local energy decomposition, where the total energy $E_{total}$ is partitioned into local atomic contributions:
\begin{equation}
    E_{total} = \sum_{i=1}^{N} \varepsilon_i (\mathcal{D}_i)
\end{equation}
where $\varepsilon_i$ is the energy of atom $i$ predicted based on its local chemical environment descriptor $\mathcal{D}_i$ within a cutoff radius $\mathcal{R}_c$.

For large-scale biomolecular systems, all atom models adopt a fragmentation-based approach to balance accuracy and scalability. The total potential energy of a protein system $E^{prot}$ can be decomposed into the internal energies of localized units and their respective long-range non-bonded interactions:
\begin{equation}
    E^{prot} = E^{prot\_units} + \sum_{i=1}^{n-1} \sum_{\substack{j=i+1 \\ i \in A, j \notin A}}^{n} E_{ij}^{Coulomb} + \sum_{i=1}^{n-1} \sum_{\substack{j=i+1 \\ i \in A, j \notin A}}^{n} E_{ij}^{VDW}
\end{equation}
MLPs focus on evaluating $E^{prot\_units}$. Correspondingly, the force $F_i$ acting on atom $i$ is derived as the negative gradient of the total energy, ensuring energy-force consistency:
\begin{equation}
    F_i^{prot} = F_i^{prot\_units} + \sum_{\substack{j=i+1 \\ j \notin A}}^{n} F_{ij}^{Coulomb} + \sum_{\substack{j=i+1 \\ j \notin A}}^{n} F_{ij}^{VDW}
\end{equation}

For complex environments such as aqueous solutions or biomembrane systems, modern MLPs go beyond local approximations to capture high-order interactions. These approaches include hybrid frameworks like AI$^2$BMD~\citep{wang2024ai2bmd}, which integrates ML with polarizable force fields; $\Delta$-learning schemes such as AQuaRef~\citep{zubatyuk2025aquaref} and SO3LR~\citep{kabylda2025so3lr} that learn quantum corrections ($E_{ML} = E_{baseline} + \Delta E$) relative to a baseline; and explicit many-body expansions like GEMS~\citep{unke2024gems}. These methodologies effectively recover many-body water-solute interactions that are typically overlooked by conventional empirical force fields. 

Instead of using MLP only for energies or forces, recent work ~\citep{bojan2026representing} treats MLP intermediate features as general-purpose embeddings of local protein environments, forming a structured manifold of local environments.

\paragraph{Machine Learning Potentials in Exploring Molecular Dynamics.}
MLPs have rapidly evolved from modeling small organic molecules to tackling full-scale biological systems. Early general-purpose models like ANI-1~\citep{smith2017ani} reached the accuracy of the underlying quantum mechanical reference data, while equivariant architectures such as TorchMD-NET~\citep{tholke2022equivariant} and NequIP~\citep{batzner20223} later revolutionized data efficiency and physical fidelity. As the field expands to all-atom biomolecular dynamics, diverse paradigms have emerged to handle system complexity. Espaloma~\citep{wang2022espaloma, takaba2024espaloma} takes a unique approach by predicting parameters for traditional molecular mechanics force fields rather than directly regressing energy, offering a differentiable bridge to legacy systems. In contrast, AI$^2$BMD~\citep{wang2024ai2bmd} stands out as a pioneering and accurate direct-energy framework, utilizing a fragmentation-based ViSNet architecture to achieve quantum-level precision. Addressing the complexity of solvation, GEMS~\citep{unke2024gems} employs a dual ``bottom-up'' and ``top-down'' training strategy to capture long-range many-body effects, while AQuaRef~\citep{zubatyuk2025aquaref} specializes in the high-resolution quantum refinement of experimental structures. Finally, SO3LR~\citep{kabylda2025so3lr} emphasizes physical integration and extreme scalability, enabling stable simulations of explicit solvent and complex biosystems exceeding 200,000 atoms. Together, these advancements signify a computational method shift toward versatile, high-fidelity simulations, combining the physical rigor of QM with the computational throughput essential for capturing the complex dynamics of large-scale biological systems.

Recent advances have integrated MLPs into ML/MM frameworks to capture the thermodynamics of enzymatic reactions with ab initio accuracy. To mitigate the computational cost of high-level reference data, $\Delta$-learning schemes have been widely adopted; for instance, Thodika et al. demonstrated that a $\Delta$-MLP model learning the correction between semi-empirical and DFT Hamiltonians exhibits strong transferability across different DHFR mutants and environments~\citep{thodika2025deltammml}. Addressing the boundary methods and embedding interactions is another improvement: Sha et al. utilized a reweighting mechanical embedding scheme with pseudobonds, correcting for polarization via thermodynamic perturbation~\citep{sha2025mmml}, while Gradisteanu et al. developed an electrostatic machine learning embedding to explicitly model static and induced environmental effects~\citep{gradisteanu2025mmml}. Furthermore, Wang et al. implemented a robust link-atom boundary scheme combined with metadynamics to explore stereoselectivity in Diels-Alderases~\citep{wang2025mmmlmetad}. Despite these methodological strides, most current models focus on simple one-step reactions, like D-A reaction and rearrangement reaction, rather than complex multi-step catalytic cycles.

\paragraph{Tackling Key Challenges: Long Range Interaction, Generalization and Computation Cost.}
Despite their promise, applying MLPs to complex biological systems faces significant hurdles, particularly regarding the locality assumption of standard GNNs. Because GNNs typically rely on message passing within a fixed cutoff radius, they struggle to capture long-range physical dependencies, such as electrostatics and dispersion, within a limited number of layers. To overcome this, recent frameworks like AI$^2$BMD and SO3LR explicitly incorporate classical physical terms or global attention mechanisms to recover these non-local interactions with a favorable balance between accuracy and throughput. Furthermore, achieving robust generalization across the vast chemical space remains a bottleneck; while models excel at standard amino acids, they must also accommodate heterogeneous components such as drug-like ligands, cofactors, and metal ions---a challenge currently addressed through massive, diverse training datasets (e.g., SPICE~\citep{eastman2023spice},~\citep{yang2024mattersim}) and transfer learning. Rollout stability under distribution shift is another practical concern, since small force errors can accumulate over long simulations; GGND~\citep{hong2026geometric} addresses this with a plug-and-play diffusion-style refinement module that improves extrapolation and stability in long MD rollouts. Finally, the computational overhead of MLPs relative to classical force fields continues to limit access to microsecond-scale dynamics, driving the need for optimized inference architectures and enhanced sampling strategies to bridge the timescale gap.

\subsection{Machine Learning Coarse Grained Models}\label{subsec:mlcg}

While all-atom MLPs achieve quantum-level precision by explicitly modeling every atomic interaction, they remain fundamentally constrained by the ``curse of dimensionality'' and the vast separation of timescales in biological processes. Even with high-performance architectures like SO3LR, simulating large-scale phenomena such as protein folding or multi-protein assembly requires sampling billions of time steps—a feat that remains computationally prohibitive when every atom is explicitly propagated. This necessitates a shift from high-fidelity force approximation to high-efficiency manifold learning. Coarse-grained (CG) models address this by strategically partitioning the system into simplified ``beads'', effectively smoothing the rugged energy landscape and allowing for much larger integration time steps. By leveraging machine learning to learn the many-body potential of mean force, these models aim to retain the essential thermodynamics and kinetics of the all-atom ensemble while operating at the massive spatial and temporal scales required for biological discovery.

\paragraph{Problem Formulation.}
The central objective of CG modeling is to reduce the dimensionality of a biomolecular system by mapping high-resolution all-atom coordinates $\mathbf{r}$ to a reduced set of ``beads'' $\mathbf{R}$ via a mapping operator $\Xi(\mathbf{r}) = \mathbf{R}$. The challenge lies in learning a CG potential energy function, $U_{CG}(\mathbf{R})$, that effectively reproduces the thermodynamics of the original system. Mathematically, this corresponds to learning the many-body potential of mean force, defined as the free energy of the coarse-grained variables:
\begin{equation}
U_{CG}(\mathbf{R}) = -k_B T \ln \left( \int e^{-\beta U_{AA}(\mathbf{r})} \delta(\Xi(\mathbf{r}) - \mathbf{R}) d\mathbf{r} \right) + C
\end{equation} 
where $U_{AA}$ is the all-atom potential and $\beta = 1/k_B T$.

\paragraph{Motivation.}
Classical CG force fields like Martini ~\citep{souza2021martini} typically rely on predefined, pairwise functional forms (e.g., harmonic bonds, Lennard-Jones potential) that often fail to capture the complex, entropic, and multi-body effects arising from the renormalization of atomic degrees of freedom. Machine learning offers a solution by approximating the complex free energy surface without rigid functional constraints. This enables the capture of accurate folding thermodynamics and kinetics at orders of magnitude lower computational cost than all-atom MD, bridging the gap between accuracy and timescale.

\paragraph{Machine Learning Coarse Grained Models in Protein Dynamics.}
Early deep learning approaches focused on direct force matching. CGnets~\citep{wang2019cgnet} pioneered this by using fully connected networks with prior energy regularization to learn multibody interactions from all-atom forces, though they were restricted to fixed topologies. To address transferability, CGSchNet~\citep{husic2020cgschnet, charron2025cgschnet} and TorchMD-Net~\citep{tholke2022equivariant,majewski2023torchnet} integrated GNNs and equivariant transformers, respectively. By learning features from local chemical environments rather than global geometries, these models allow force fields trained on small peptides to be transferred to unseen protein sequences and larger topologies.

Moving beyond supervised force regression, recent methods leverage generative and variational paradigms. PCCG~\citep{ding2022pccg} reformulates force field learning as a classification problem using Noise Contrastive Estimation with umbrella sampling, enabling efficient parameterization without expensive force labels. Similarly, Flow-Matching~\citep{Kohler2023flowmatching} utilizes normalizing flows to minimize relative entropy, achieving high data efficiency solely from structural distributions.

The most recent wave of innovation integrates deep learning with physical constraints or interpretable interactions. 2-for-1-diffusion~\citep{arts20232for1diffusion} demonstrates that a single score-based model can serve dual purposes: generating i.i.d. equilibrium samples and driving dynamics via the learned score function ($ \nabla \log p(\mathbf{R}) \approx -\nabla U(\mathbf{R})$). Building on this, ScoreMD~\citep{plainer2025consistent} addresses the physical inconsistency between the denoising distribution and the energy landscape by introducing Fokker-Planck regularization, ensuring the model functions as both a high-quality generator and a stable, conservative force field. HFM~\citep{ripken2026learninghamiltonianflowmaps} learns Hamiltonian flow maps from instantaneous samples to predict long-time dynamics with improved stability and approximate structure preservation. Complementary to these dynamics models, InvertibleCG~\citep{chennakesavalu2023thermodynamicconsistency} optimizes a state-dependent invertible map to ensure thermodynamic consistency during back-mapping, while GNN-LRP~\citep{bonneau2025gnnlrp} applies Layer-wise Relevance Propagation to decompose opaque GNN potentials into interpretable physical interactions.

\subsection{Machine Learning Collective Variables}\label{subsec:mlcv}

The identification of optimal collective variables (CVs) is a prerequisite for efficient enhanced sampling. Machine learning  has revolutionized this field by automating the discovery of CVs through three primary paradigms: dimensionality reduction, reinforcement learning, and generative modeling.

\paragraph{Dimensionality Reduction and Kinetic Learning.}

A foundational approach involves projecting high-dimensional atomic coordinates into low-dimensional latent spaces that capture significant conformational changes. Early supervised methods like SML~\citep{sultan2018sml} utilized classification algorithms to discriminate between metastable states. Unsupervised approaches, particularly autoencoders, were popularized by MESA~\citep{chen2018mesa, chen2018mesa2}, which drives sampling along the nonlinear manifold learned from MD trajectories. Subsequent frameworks such as DESP~\citep{salawu2021desp} and FEBILAE~\citep{belkacemi2022febilae} further refined autoencoder-based biasing to explore free energy landscapes. Beyond geometric compression, recent methods incorporate kinetic information to identify slow degrees of freedom. DeepTICA~\citep{luigi2021deeptica} employs neural networks to approximate the eigenfunctions of the transfer operator for variational approach, extracting slow modes directly from biased simulations like OPES. Similarly, MaxEnt-VAMPNet~\citep{kleiman2023maxentvampnets} and TLC~\citep{park2025tlc} leverage maximum entropy and time-lagged correlations to resolve kinetic barriers. Other notable dimensionality reduction techniques include LDA~\citep{sasmal2023lda}, which applies linear discriminant analysis for CV discovery, and Deep-TDA~\citep{trizio2021enhanced}, which trains a feed-forward neural network with a discrimination criterion on symmetry-invariant physical descriptors (e.g., interatomic distances, angles) collected from short unbiased simulations of metastable basins, projecting them into a low-dimensional space where each basin follows a preassigned distribution, thereby enabling efficient enhanced sampling with fewer CVs even in multistep chemical processes. DynaProt~\citep{bafna2026learning} aims for a lightweight alternative to full MD by reducing CV dimension to support residue-level dynamics through multivariate Gaussians at two scales: per-residue covariance matrices and scalar pairwise covariances, directly from a single static structure. 

\paragraph{Reinforcement Learning and Adaptive Sampling.}

To actively navigate rugged energy landscapes, RL-based frameworks treat conformational sampling as a decision-making process. REAP~\citep{shamsi2018reap} and its multi-agent extension MA-REAP~\citep{kleiman2022mareap} select order parameters to maximize exploration rewards. DeepDriveMD~\citep{lee2019deepdrivemd} provides a scalable infrastructure for such adaptive loops. More recent strategies reformulate the exploration problem: AdaptiveBandit~\citep{perez2020adaptivebandit} treats MSM states as arms in a multi-armed bandit problem to optimize restarting points, while Adaptive CVGen~\citep{shen2024adaptivecvgen} dynamically updates CV weights using dual RL agents to balance exploration and exploitation. Similarly, policy driven adaptive SMD~\citep{ho2022policydrivenadaptivesmd} uses RL to steer steered molecular dynamics. Distinct from standard RL, RiD~\citep{wang2021rid, fan2025rid} operates as a concurrent active learning scheme; it utilizes an ensemble of neural networks to approximate the potential of mean force on-the-fly, using uncertainty quantification to guide the system out of local minima without pre-existing datasets.

\paragraph{Generative Models with CV Constraints.}

The latest wave of methods leverages generative priors and differentiable physics. DiffSim~\citep{sipka2023diffsim} and DeepLNE~\citep{fröhlking2024deeplne} utilize end-to-end differentiable simulations to learn bias potentials or transition paths by optimizing path-integral loss functions. To address data scarcity, Geodesic Interpolation~\citep{yang2024geoint} employs Riemannian manifold operations to synthetically augment transition paths. Furthermore, the integration of well-pretrained models like AlphaFold (AF) and BioEmu has led to hybrid protocols: AF2-RAVE~\citep{vani2023af2rave} use multiple AF predictions as initial seeds for Boltzmann-ranked ensembles recovering using deep information bottleneck objectives, while AlphaFold-Metainference~\citep{brotzakis2025alphafoldmainteinference} uses predicted distograms as restraints to model disordered regions. BioEmu-CV~\citep{parklearning} learns CVs by repurposing BioEmu for time-lagged, CV-conditioned generation, encouraging the CV to focus on slow modes. Besides learning CVs, WT-ASBS~\citep{nam2026enhancing} adds an explicit exploration bias in CV space (a repulsive term around recent samples) and uses reweighting to recover thermodynamic estimates. In parallel, diffusion-based samplers such as DDPM-REST2~\citep{benayad2025ddpmrest2} and LAST~\citep{tian2023last} combine score-based generation with thermodynamic sampling to improve exploration of configurational space.

\subsection{Towards Advanced Kinetic and Thermodynamic Modeling }

The preceding subsections addressed how machine learning can replace or augment the core components of an MD pipeline, including the potential energy surface, the level of resolution, and the choice of reaction coordinates. Building on these foundations, machine learning is now also transforming downstream tasks that consume MD output, including the identification of transition states and the acceleration of Free Energy Perturbation (FEP) calculations.

\paragraph{Learning for transition states.}

Committor functions are theoretically precise reaction coordinates related to transition pathways, but solving their high-dimensional equations has traditionally been extremely difficult. Many deep learning methods were developed to estimate committors and automatically identify transition states in complex conformational changes. Architectures like GVP-GNNs~\citep{arredondo2025gvpgnn} are now being used to predict committors directly from atomic coordinates. This grants atom-level interpretability without prior topological assumptions, automatically pinpointing key atoms in the transition. Other methods utilizing Siamese neural networks~\citep{chen2023siamese} and iterative variational learning~\citep{megías2026ivl} can now simultaneously learn the committor and its associated transition string, providing a comprehensive view of the reaction mechanism. TS-DAR~\citep{liu2025tsdar} innovatively treats transition states as out-of-distribution data in the potential space of the hypersphere. By introducing a regularized loss function of dispersion and variational principle, the method breaks through the previous limitation of finding transition states between two known states, and can automatically and simultaneously identify all unknown transition states of proteins from MD trajectories.

\paragraph{Learning for FEP.}

FEP is the gold standard for predicting binding affinities in structure-based drug design. However, its high computational cost and complex protocol setup traditionally limit its throughput. The latest end-to-end biomolecular foundation models such as Boltz-2~\citep{passaro2025boltz} are reshaping this status quo. Boltz-2 is among the first open-source AI models to approach the accuracy of rigorous physical FEP in predicting small molecule-protein binding affinity, while being orders of magnitude faster in inference~\citep{passaro2025boltz}. The Boltz-ABFE~\citep{thaler2026boltzabfe} method directly uses the high-quality 3D structure of the protein-ligand complex predicted by Boltz-2 as the initial conformation to initialize the MD simulation of absolute free energy perturbation, which completely expands the application boundary of FEP technology in the amorphous structure scenario. In addition to providing high-quality initial structures, ML is also used for automatic error correction and sampling optimization. Platforms such as FEP $\Omega$~\citep{giannakoulias2025fep} combine with downstream ML for error correction for sparse experimental labels. At the same time, the combination of active learning and FEP~\citep{pinxteren2025fepworkflows} can be simulated by intelligently selecting the most promising ligands, greatly improving the virtual screening efficiency in massive chemical spaces.

\section{Data}\label{sec:data}

% table + url column

\begin{table}[htbp]
    \centering
    \scriptsize
    \caption{Overview of public protein datasets: source, scale, and system types. Note that some databases such as PDB are still being updated.}
    \label{tab:datasets}
    \begin{tabularx}{\textwidth}{@{}l l X l@{}}
        \toprule[1.5pt]
        \textbf{Dataset} & \textbf{Source} & \textbf{Scale} & \textbf{Systems} \\
        \midrule[1pt]

\multicolumn{4}{c}{\textbf{Static Structures}} \\
\midrule[1pt]
PDB~\citep{berman2000pdb} & Experimental & $\sim$215K structures & Monomer, Complex \\
AFDB~\citep{varadi2024alphafold} & Distillation & 214M structures & Monomer \\
AFESM~\citep{yeo2025afesm} & Distillation  & 821M structures & Monomer (Metagenomic) \\
CDDB~\citep{reidenbach2025consistent} & Distillation & 455k structures & Monomer \\
Teddymer~\citep{didi2026proteinacomplexa} & Distillation & 510k clusters & Dimer\\
\midrule[1pt]
% MD Simulations
\multicolumn{4}{c}{\textbf{MD Simulation Trajectories}} \\
\midrule
ATLAS~\citep{vander2024atlas} & Simulation (MD) & 1,390 proteins (100 ns) & Protein Monomer \\
DynamicPDB~\citep{liu2024dynamicpdb} & Simulation (MD) & 12,643 proteins (1 $\mu$s) & Protein Monomer \\
DynaRepo~\citep{mokhtari2025dynarepo} & Simulation (MD) & $\sim$720 systems (500 ns) & Monomer, Complex \\
mdCATH~\citep{mirarchi2024mdcath} & Simulation (MD) & 5,398 domains (up to 500 ns) & Monomer (Domains) \\
Fast-Folders~\citep{lindorff2011fast} & Simulation (MD) & 12 domains (up to ms scale) & Mini-protein \\
Shaw2010~\citep{shaw2010atomic} & Simulation (MD) & 2 proteins (ms scale) & Mini-protein \\
ProteinConformers~\citep{zhou2025proteinconformers} & Simulation (MD) & 381K conformations; 87 targets & Protein Monomer \\
MDRepo~\citep{roy2024mdrepo} & Simulation (MD) & $\sim$4.2K community trajectories & Monomer, Complex \\
GPCRmd~\citep{rodríguez-espigares2020gpcrmd} & Simulation (MD) & 570 systems ($3 \times 500$ ns) & Membrane (GPCR) \\
MemProtMD~\citep{stansfeld2015memprotmd} & Simulation (CGMD) & 2,294 proteins (1 $\mu$s) & Membrane Proteins \\
AF-CALVADOS~\citep{von2025afcalvados} & Simulation (CGMD) & 12,483 human proteins & Monomer, MDPs \\
MISATO~\citep{siebenmorgen2024misato} & Simulation (MD) & $\sim$17K structures (8 ns) & Protein-Ligand \\
DD-13M~\citep{li2025dd13m} & Simulation (MD) & 26,612 trajectories & Protein-Ligand \\
ManyPeptidesMD~\citep{tan2025prose} & Simulation (MD) & 21,700 peptides (200 ns) & Peptides (2-8 AA) \\
\midrule[1pt]
% Disordered, Functional & Other Experimental
\multicolumn{4}{c}{\textbf{Disordered Proteins \& Experimental Measurements}} \\
\midrule[1pt]
BMRB~\citep{hoch2023bmrb} & Experimental (NMR) & $>$11M chemical shifts & Proteins, Nucleic acids \\
SASBDB~\citep{kikhney2020sasbdb} & Experimental (SAS) & $>$1K entries & Monomer, Complex \\
MobiDB~\citep{piovesan2025mobidb} & Hybrid & $>$200M entries & IDPs, IDRs \\
DisProt~\citep{nugnes2025disprot} & Hybrid & 3,201 proteins & IDPs, IDRs \\
% Hybrid means experimental (curated) and predicted

\bottomrule[1.5pt]
\end{tabularx}
\end{table}

The availability of high-quality data is critical for training and benchmarking generative models and machine learning potentials for protein dynamics. We summarize the key datasets in Table~\ref{tab:datasets} and categorize them into three main groups: static structures, molecular dynamics trajectories, and disordered proteins with experimental measurements.

\paragraph{Static Structures.}
The Protein Data Bank (PDB)~\citep{berman2000pdb} serves as the foundational repository for experimentally determined 3D structures of proteins and complex assemblies. Moving beyond experimental data, massive databases of predicted structures have emerged via deep learning distillation. AlphaFold Protein Structure Database (AFDB)~\citep{jumper2021highly, varadi2024alphafold} contains over 200 million predicted structures, capturing the structural coverage of almost all cataloged proteins in UniProt~\citep{uniprot2025uniprot}. Building on this, the AFESM dataset~\citep{yeo2025afesm} expands the coverage to metagenomic sequences, providing hundreds of millions of predicted structures for previously unseen proteins. Additionally, large-scale entirely synthetic, fully atomistic de novo predicted sequences and structures have also been recently adopted to unlock vast structural diversity~\citep{reidenbach2025consistent}. Teddymer also presents more than 500k dimer clusters as a database generated by treating each domain as a chain in AFDB~\citep{didi2026proteinacomplexa}.

\textbf{Opportunities and Challenges}: \textit{In vitro} crystal structures (PDB) provide ground-truth atomic representations governed by true physical constraints, but they suffer from severe sparsity and known experimental biases (such as crystal packing artifacts). Conversely, distillation and synthetic approaches offer an unprecedented scale and broad sequence-structure coverage, unlocking vast amounts of data to pre-train large-scale generative models. However, these synthetic structures predominantly reflect the single dominant, static state favored by the underlying prediction algorithm, completely missing the ruggedness of genuine thermodynamic ensembles. Furthermore, synthetic data often harbors predictive artifacts or unphysical stereochemistry. Consequently, applying static structures to learn protein dynamics necessitates rigorous filtering, e.g., culling low-confidence predictions via pLDDT scores, ensuring sufficient secondary structures, or enforce consistency~\citep{reidenbach2025consistent}, so that models learn physically meaningful relationships rather than overfitting to the unphysical biases inherent in modern structural predictors.

\paragraph{MD Simulation Trajectories.}
% general proteins
While static structures offer single snapshots, characterizing protein dynamics requires continuous trajectories. Existing trajectory datasets can be broadly categorized by their target systems and timescales. For \textbf{globular monomeric proteins}, datasets like ATLAS~\citep{vander2024atlas} and DynamicPDB~\citep{liu2024dynamicpdb} provide extensive 100~ns to 1~$\mu$s MD simulations for thousands of targets, while the mdCATH dataset~\citep{mirarchi2024mdcath} features simulation data at various temperatures for domains. To study \textbf{long-timescale folding events} and conformational diversity, the fast-folding proteins dataset~\citep{lindorff2011fast}, the pioneering millisecond-scale trajectories of BPTI (Shaw2010)~\citep{shaw2010atomic}, and ProteinConformers~\citep{zhou2025proteinconformers} remain critical benchmarks. Beyond standard monomeric proteins, several \textbf{specialized datasets} focus on complex interactions and environments. For protein-ligand interactions, MISATO~\citep{siebenmorgen2024misato} and DD-13M~\citep{li2025dd13m} supply valuable QM/MM and unbinding MD trajectories. Membrane proteins and GPCRs are extensively covered by MemProtMD~\citep{stansfeld2015memprotmd} and GPCRmd~\citep{rodríguez-espigares2020gpcrmd}, respectively. Finally, for \textbf{bottom-up and scalable ML models}, ManyPeptidesMD~\citep{tan2025prose} provides dense sampling of small peptide dynamics, and AF-CALVADOS~\citep{von2025afcalvados} provides coarse-grained MD simulations for large human proteins. Furthermore, community-curated platforms like DynaRepo~\citep{mokhtari2025dynarepo}, MDRepo~\citep{roy2024mdrepo}, and MDDB~\citep{amaro2025mddb} aggregate and standardize massive simulation records for machine learning.

\textbf{Opportunities and Challenges}: MD trajectories capture the time-correlated kinetic behaviors and thermodynamic distribution that static data lacks. Learning from this data gives models the ability to recover transition probabilities, reactive pathways, and sequence-dependent flexibility. However, generating large-scale, diverse, and heavily sampled MD datasets is remarkably expensive. Crucially, the accuracy and reliability of these datasets are fundamentally tied to the quality of the underlying empirical force fields used during simulation, directly impacting the validity of the underlying dynamics landscape explored. Datasets curated from classical simulations inherently carry these modeling inaccuracies. Furthermore, short simulations can easily become trapped in local energy minima, leading to highly auto-correlated distributions that fail to properly represent the global equilibrium. When generative methods rely solely on this data, they are vulnerable to overfitting localized phase spaces and perfectly reproducing the imperfections of the molecular mechanism force fields, highlighting the need for higher-fidelity quantum data or enhanced sampling strategies.

\paragraph{Intrinsically Disordered Proteins and Experimental Measurements.}
Intrinsically Disordered Proteins (IDPs) and Regions (IDRs) present unique challenges due to their highly dynamic, ensemble-like nature. Databases such as MobiDB~\citep{piovesan2023mobidb, piovesan2025mobidb} and DisProt~\citep{nugnes2025disprot} curate experimental annotations and predictions of disorder, serving as critical references for models predicting IDP ensembles. Additionally, experimental measurements of dynamics and flexibility, such as NMR chemical shifts from BMRB~\citep{hoch2023bmrb} and small-angle scattering profiles from SASBDB~\citep{kikhney2020sasbdb}, provide macroscopic or time-averaged thermodynamic constraints that can steer or validate generative ensemble models.

\textbf{Opportunities and Challenges}: Such ensemble-averaged measurements offer the unique advantage of reflecting flexible macro-states observed under physiological conditions, rather than static artificial snapshots or computationally biased simulations. For AI models, substituting or supplementing force field learning with direct observational signals can correct severe kinetic inaccuracies and guide the sampling towards experimentally validated free-energy basins. This is particularly vital for modeling IDPs, where no true "gold-standard" classical force field yet exists to reliably simulate their highly fluxional ensembles. This distinct lack of a universally accurate physical model provides a profound opportunity for data-driven generative models to learn directly from structural ensembles and experimental macro-signals, bypassing empirical force field bottlenecks altogether. However, the central computational challenge remains: experimental constraints are intrinsically sparse, noisy, and indirect. NMR or SAXS profiles do not provide explicit 3D atomic coordinates. Instead, they entail a mathematically complex, one-to-many relationship where numerous distinct structural ensembles could theoretically reproduce the same average signal. Bridging generative models with experimental constraints thus requires robust, differentiable forward models capable of computing expectation values back to atomic representation, a task complicated by substantial computational overhead and empirical approximations within the forward models themselves.

\section{Discussion}\label{sec:discussion}

The intersection of generative AI and molecular dynamics has catalyzed a paradigm shift in how we model and understand protein dynamics. Throughout this survey, we have explored the rapid evolution from traditional physics-based simulations to sophisticated machine learning frameworks. As detailed in the preceding sections, this progress stems from three primary pillars: learning from vast structural datasets to sample conformational ensembles and simulate trajectories (Section~\ref{sec:structural_data}), leveraging physical energy signals to steer generative priors toward thermodynamic fidelity (Section~\ref{sec:energy_signals}), and integrating machine learning deeply into MD pipelines via neural potentials, coarse-graining, and learned collective variables (Section~\ref{sec:learn_mds}). Supported by the growing availability of both simulated and experimental datasets (Section~\ref{sec:data}), these methodologies effectively bridge the critical gap between static atomic snapshots and true, time-aware biomolecular motions.

As these foundation models mature, their significance extends far beyond computational biophysics into transformative real-world applications across life sciences. In structure-based drug discovery, characterizing the full conformational ensemble enables the reliable identification of cryptic pockets and elusive allosteric sites that are completely invisible in static crystal structures. Furthermore, generative conformational sampling combined with ML-based potentials may enable faster binding free-energy estimation and improve the practical utility of virtual screening, although achieving reliable chemical accuracy remains system- and protocol-dependent.
% now allows for rapid absolute binding free energy calculations, pushing virtual screening closer to chemical accuracy at a fraction of the traditional computational cost. 

Targeted generative models, such as PLACER~\citep{ivan2025placer}, are already successfully capturing the dynamic behavior of ligands within protein binding sites. In the domain of protein and enzyme engineering, the ability to accurately trace transition states and reaction coordinates empowers the design of highly efficient \textit{de novo} biocatalysts and dynamic nanosensors, moving the field beyond the engineering of rigid scaffolds toward the programmable design of biological function.

Despite these remarkable achievements, the full potential of AI in modeling protein dynamics is hindered by several critical bottlenecks. To overcome these limitations, the research community must actively address the following interconnected \textbf{future directions}:

\begin{itemize}
    \item \textbf{Generative Force Fields:} Traditional Machine Learning Potentials (MLPs, Section~\ref{sec:learn_mds}) act as neural distillations of Density Functional Theory (DFT) or classical Molecular Mechanics Force Fields (MMFFs). While highly efficient, this supervised approach is inherently bottlenecked by the biases, inaccuracies, and simplifications of the underlying empirical or quantum labels. A major future opportunity lies in developing \textbf{Generative Force Fields}: models that establish the energy landscape directly from the data distribution of structural ensembles without relying on potentially biased energy labels. Ensuring that these generative priors naturally entail conservative force fields while adhering to detailed balance and thermodynamic consistency remains a largely unsolved algorithmic challenge, but presents a pathway to bypass legacy label biases entirely. A key challenge is that such data-derived landscapes may not be uniquely identifiable without physical constraints or experimental calibration.

    \item \textbf{Integration of Sparse Experimental Constraints:} As discussed in Section~\ref{sec:data}, most current ML models still rely heavily on computational MD trajectories or algorithmically distilled synthetic structures, making them vulnerable to empirical modeling biases or unphysical prediction artifacts. A major future direction involves directly incorporating heterogeneous, sparse experimental data—such as NMR chemical shifts, SAXS profiles, and cryo-EM equilibrium densities—as foundational training signals or for rigorous post-training \textbf{calibration}. Using these highly accurate experimental signals to calibrate generated ensembles will critically correct localized kinetic inaccuracies, grounding generative sampling directly in true physiological states \textit{in vivo}. Bridging this gap will require robust differentiable forward models capable of computing expectation values from atomic coordinates back into macroscopic observables.

    \item \textbf{Biomolecular Foundation Models:} The development of unified computational architectures marks a critical frontier. By coherently integrating diverse modalities of data, encompassing organic molecules, protein and nucleotide sequences, evolutionary profiles, static structural snapshots, and continuous simulation trajectories, future foundation models (e.g., Boltz-2~\citep{passaro2025boltz}) hold the potential for robust zero-shot generalization across disparate dynamic tasks, mirroring the profound impact of foundation models in natural language processing.

    \item \textbf{Scaling Boltzmann Generators:} As highlighted in Section~\ref{sec:energy_signals}, applying exact Boltzmann generators to large protein complexes remains exceptionally challenging due to the curse of dimensionality and the high computational cost of the likelihood evaluations required for unbiased reweighting. These difficulties become even more pronounced for protein--ligand systems and other heterogeneous biomolecular assemblies, where binding-mode diversity, interfacial rearrangements, and induced-fit or allosteric couplings substantially enlarge the relevant conformational landscape. Future efforts must therefore focus on architectural innovations that scale these systems (e.g., via efficiently parallelized neural architectures or diffusion-based amortized samplers) while maintaining tractable accept-reject rates during importance sampling.

    \item \textbf{Trajectory-based Benchmarking:} For methods generating continuous time-series trajectories, such as autoregressive steps or Markovian transitions, error accumulation over extended horizons severely restricts models from simulating microsecond phenomena without unphysical structural branching. This methodological hurdle underscores a conspicuous lack of standardized, comprehensive community benchmarks. Establishing rigorous evaluation frameworks to measure the temporal fidelity, kinetic consistency, and long-term stability of generative trajectories is an urgent necessity for standardizing progress in the field.

    \item \textbf{Uncertainty Quantification:} If a protein dynamics model can produce calibrated uncertainty estimates alongside its predicted updates, these signals could indicate when a rollout is leaving poorly explored regions of conformational space and motivate uncertainty-aware adaptive resampling; more generally, limiting rollout length based on model confidence is a standard way to reduce compounding error in sequential prediction~\citep{Imbalzano2021}. While AlphaFold's pLDDT excels at static, per-residue scores~\citep{jumper2021highly}, we are interested in a calibrated uncertainty over rollout dynamics. 
\end{itemize}

\textbf{Closing Remarks.} In summary, generative models has shown great potential in decoupling the sampling of biological phase space from the extreme computational cost of classical Newtonian integration. We are entering a new era of structural biology where the ability to rapidly and accurately simulate protein dynamics will soon rival our current capacity to accurately predict static structures. By increasingly unifying data-driven generative architectures with fundamental physical laws and rigorous experimental calibration, the research community will witness many advances toward an ultimate and transformative goal of the comprehensive, predictive understanding of biomolecular function in motion.

\section*{Author Contribution} 
H.T., L.S., and Y.Z. mainly conducted the literature review and wrote Chapter 2, Chapter 3, and Chapter 4, respectively. X.L. contributed to the revision of the manuscript and wrote part of the Chapter 4. J.T. provided overall supervision, reviewed and revised the manuscript. J.L. conceived the structure of the survey, directed the project, conducted the literature review, and wrote the introduction, data \& discussion sections. All authors read and approved the final manuscript.

% ===================================

% bibliography
% \printbibliography
\bibliography{refs}
\end{document}

%% file: math.tex
\usepackage{amsmath,amsfonts,bm}

\def\eqref#1{equation~\ref{#1}}
\def\1{\bm{1}}

\def\rva{{\mathbf{a}}}

\def\rvx{{\mathbf{x}}}
\def\rvy{{\mathbf{y}}}
\def\rvz{{\mathbf{z}}}

\def\rmX{{\mathbf{X}}}

\DeclareMathAlphabet{\mathsfit}{\encodingdefault}{\sfdefault}{m}{sl}
\SetMathAlphabet{\mathsfit}{bold}{\encodingdefault}{\sfdefault}{bx}{n}

\newcommand{\calvar}[1]{\ensuremath{\mathcal{#1}}}

\newcommand{\calN}{\calvar{N}}